\documentclass[a4paper,11pt]{article}
\pdfoutput=1 

\usepackage{jcappub} 

\usepackage[T1]{fontenc} 
\usepackage{mathrsfs}
\usepackage{multirow}
\usepackage{algorithm} 
\usepackage{algpseudocode}
\usepackage{lipsum}
\usepackage{longtable,tabu}
\usepackage[normalem]{ulem}
\usepackage{hyperref}
\usepackage{natbib}
\title{The Neutron Star Outer Crust Equation of State: A Machine Learning approach}

\author[a]{Utsav Anil Murarka, }
\author[a]{Kinjal Banerjee, }
\author[b,1]{Tuhin Malik, \note{Corresponding author.}}
\author[b]{Constan\c ca Provid\^encia}

\affiliation[a]{BITS-Pilani, Department of Physics, K.K. Birla Goa Campus, GOA - 403726, India}
\affiliation[b]{CFisUC, 
	Department of Physics, University of Coimbra, P-3004 - 516  Coimbra, Portugal}

\emailAdd{utsavm1804@gmail.com}
\emailAdd{tuhin.malik@uc.pt}
\emailAdd{kinjalb@gmail.com}
\emailAdd{cp@fis.uc.pt}

\abstract{Constructing the outer crust of the neutron stars requires the knowledge of the Binding Energy (BE) of the atomic nuclei. Although the BE of a lot of the nuclei is experimentally determined and can be obtained from the AME data table, for the others we need to depend on theoretical models. There exist a lot of physical theories to predict the BE, each with its own strengths and weaknesses. In this paper we apply Machine Learning (ML) algorithms on AME2016 data set to predict the Binding Energy {of atomic nuclei}. The novel feature of our work is that it is model independent. We do not assume or use any nuclear physics model but use only ML algorithms directly on the AME2016 data set. Our results are further refined by using another ML algorithm to train the errors of the first algorithm, and repeating this process iteratively. Our best algorithm gives $\sigma_{\rm rms} \approx 0.58$ MeV for Binding Energy on randomized testing sets. This is comparable to all physics models or ML improved physics models studied in literature till date.  Using the predictions of our Machine Learning algorithm, we construct the outer crust equation of state (EoS) of a neutron star and show that our model is comparable to existing models. This work also demonstrates the use of various ML algorithms and a detailed analysis on how we arrived at our best algorithm. It will help the physics community in understanding how to choose an ML algorithm which would be suited for their data set. Our algorithms and best fit model is also made publicly available for the use of the community.}

\begin{document}
\maketitle
\flushbottom
\section{Introduction}
The Neutron Stars (NSs) are the ideal cosmic laboratories for the unique conditions and phenomena that go beyond the physics standard scenario. The NS interior consists of three main regions, namely, an outer crust, an inner crust, and a core. The NS core is incredibly dense and reaching densities up to few times the nuclear saturation density $\rho_0$ ($\rho_0 \sim 2.7 \times 10^{14}$ gm/cm$^{3}$).
In this paper we confine our interest to the physics of the neutron star outer crust. The densities of the outer crust of neutron star ranges from $10^{4} \mathrm{~gm} / \mathrm{cm}^{3}$ to a neutron drip density of about $4 \times 10^{11} \mathrm{~gm} / \mathrm{cm}^{3}$. At such densities most energetically favorable states for nucleons is to cluster into individual nuclei and arrange themselves in a solid body-centred cubic (bcc) lattice in order to minimize the Coulomb repulsion and to stabilize against $\beta$ decay by the surrounding electron gas \cite{Baym1971}. The electrons are present to maintain the charge neutrality but they are not bound to nuclei and move freely throughout the crust. The composition of the NS outer crust can be determined by minimizing the the Gibbs free energy or chemical potential at a certain pressure with respect to the atomic mass number A and the atomic charge Z. So, the knowledge of nucleus binding energy for different isotopes and isobars are essential. It is to be notated, an atom denoted by $^{A}_{Z} {\rm X}$, has mass number $A=(N+Z)$, where $N$ is the number of neutrons, and $Z$ is the number of protons. The BE of a nucleus is defined as, $BE(Z, N) \equiv Z m_{p}+N m_{n}-M(Z, N)$, where $m_p$, $m_n$ and $M(Z,N)$ are the individual mass of proton, neutron and total mass of the nucleus, respectively \cite{Krane1988}.

At the top of the outer crust where density is low, the electronic contribution to the chemical potential is negligible, so the Coulomb lattice is populated by stable $\mathrm{Fe}$ or $\mathrm{Ni}$ nuclei and the knowledge of nuclear binding energy are very well known here. The number of electrons is the same as protons. As density increases, the electronic contribution becomes significant and the electrons get captured on the protons. When the proton fraction becomes low enough and the symmetry energy large, it become energetically favorable for the system to jump into the $N = 50$ region. Finally, bottom layers of the outer crust are in the $N = 82$ region which requires knowledge of nuclear masses in the region from $^{132}{\rm Sn}$ ($Z = 50$) all the way down to $^{118}{\rm Kr}$ ($Z = 36$). In this region the theoretical extrapolations are unavoidable as almost no experimental information. So, the composition of the NS outer crust is highly sensitive to nuclear BE and leads to a close connection between astrophysics and nuclear physics \cite{Baym1971,RocaMaza2008,Chamel2008,bertulani2012neutron}(and references therein).
In fact, the BE is one of the fundamental properties of atomic nuclei and most of the other properties of atomic nuclei like mass, decay lifetimes and reaction rates are governed largely by the BE. It also plays a significant role in various nuclear structure information, such as nuclear pairing correlation, shell effect, deformation transition, and so on \cite{Lunney2003}. The BE is also used widely to constrain the parameters of the theory of nuclear effective interactions \cite{Bender2003}.

In recent years, the experimental measurements of nuclear BE have achieved a great success, in the atomic mass evaluation AME2016 \cite{Wang2017},
3435 nuclei have been measured in the laboratories around the world. However, several of them ($\approx
25\%$) are not strictly determined experimentally. 
The BE of these nuclei has been measured by the trends from the mass surface (TMS) defined by the neighboring nuclei. 
It has been observed experimentally the atomic mass forms a surface
when it is displayed as a function of N and Z and due to pairing energy of nuclei, this surface can be divided into four
sheets. These mass sheets are very regular in all places unless there are changes in nuclear structure in a particular region of the
surface. This regularity in the mass surface is one of the basic properties and is employed to obtain unknown, poorly
known, or questionable masses by extrapolation from the well-known nearby mass values on the same sheet.

Many nuclei of astrophysical relevance still remain beyond the experimental reach 
\cite{Mumpower2016,RocaMaza2008,Utama2016}. Thus, theoretical modeling of nuclear theory that extrapolate BE into unknown regions of the nuclear chart becomes very important. Unfortunately, theoretical modeling of nuclei to predict BE  is challenging due to the uncertain theories of nuclear interaction
and difficulties in the quantum many-body calculations\cite{Moller1994,Tajima2001,Audi2003}. 
The first mass formula for predicting BE of a atomic nuclei is the Bethe–Weizs$\ddot{\rm a}$cker (BW) mass formula \cite{Bethe1936,Weizsacker1935}. 
This model was based on macroscopic considerations which assumes that the nucleus is a charged liquid drop. It does not account any 
microscopic effect, such as shell effect. Later, the macroscopic–microscopic models were developed by taking into account the 
microscopic effects, such as the finite-range droplet model (FRDM) \cite{Moller2012} and 
the Weizs$\ddot{\rm a}$cker-Skyrme (WS) model \cite{Wang2014}. There also exists microscopic models which are 
completely based on Density Functional Theory (DFT). 
There are two types of DFT calculations, one type is based on a non-relativistic framework and other is on a relativistic framework. 
The series of Hartree–Fock–Bogoliubov (HFB) DFT models on non-relativistic framework are constructed with the Skyrme \cite{Goriely2009a,Goriely2016} or Gogny \cite{Goriely2009} effective interactions. 
More recently, the relativistic mean-field (RMF) models have been of great interest as they have been able to successfully describe various nuclear 
and astrophysical phenomena \cite{Meng2006a,Vretenar2005,Meng2006,Liang2008,Niu2013c,Niu2017,Sun2008,Niu2009,Niu2013a}. 
However, the prediction of BE of these models differs from experimentally observed values  by a Root-Mean-Squared-Error 
$\sigma_{\rm rms}\approx 3$ MeV for BW model \cite{Audi2012,Wang2012} to $\sigma_{\rm rms}\approx 0.3$ MeV for WS model \cite{Wang2014}. 
Moreover, the error is not uniform over all mass ranges. The predictions are very divergent (even up to tens of MeV) for 
lighter nuclei $A<16$ and also when extrapolated for heavy nuclei which have a large proton neutron number asymmetry.   
The accuracies of these models are not sufficient for studying excited nuclear states or for astrophysical applications like
constructing crusts of neutron stars. Therefore, there is a lot of room for improvement in the accuracy of BE prediction for 
nuclei in any mass range and also for any exited states. One of the ways of improving the predictions of these models is by using Radial Basis Functions(RBF) \cite{Wang2011,Niu2016,Niu2018b}.

In recent years, Machine Learning (ML) algorithms are being widely used in fundamental research, and physics is no exception for converting information into knowledge (see \cite{George2017,Guest2018} among others).
Machine Learning provides a powerful tool to classify and to predict patterns, even in complex data sets. The historical overview of the development of the field can be found in Refs. \cite{Lecun2015,Schmidhuber2015} and the 
recent introduction to machine learning for physicists in Refs. \cite{Carleo2019,Mehta2018}. In the area of nuclear physics  ML algorithms 
have been used to predict other nuclear properties like beta decay  \cite{Niu2019,Costiris2009}, alpha decay \cite{Rodriguez2019},
 uncertainty quantification in nuclear shell model \cite{Yoshida2018}, excited states in nuclei \cite{Regnier2019} among others. In fact, In this paper, we will concentrate on the BE prediction where learning algorithms (neural networks) was applied in 1992 \cite{Gazula1992} followed
by a series of works which further developed the predictive accuracy \cite{Gernoth1993,Athanassopoulos2004,Zhang2017}. 
More recently, further improvement has been done by employing Bayesian Neural Network (BNN) on predictions of nuclear
masses \cite{Utama2016,Niu2018} and nuclear charge radii \cite{Utama2016a}. However, all these previous works to predict nuclear masses were not completely based on learning algorithms. 
They were employed on top of a base physics model and were used only to improve the accuracy of that model.

In this work we construct the  Neutron star outer crust EoS using the prediction of nuclear BE of the Machine Learning (ML) algorithm. Although our predictions of BE via ML algorithms has been motivated from the past works cited above, it is novel and unique in the sense that we have developed an algorithm for nuclear mass/BE prediction in a model independent way without help of any nuclear physics models of BE but by only  using ML algorithms on experimental data of the 3435 nuclei given in  AME2016 \cite{Wang2017}. The only physics input we use is the nucleus is characterized that the number of protons $Z$ and  the number of neutrons $N$ as well as the TMS technique used to obtain some of the BE values in AME2016. The interesting feature of our work is that, even without having any further \textit{physics input}, our prediction for BE is comparable to all theoretical models as well as those models where ML algorithms have been used on top of a base physics model. Further, unlike the above mentioned models, the $\sigma_{\rm rms}$ of our model is low even for light and for heavy nuclei. Since the use of ML is still new in Physics, another aspect of our work is pedagogical where we have indicated a step by step procedure in choosing the optimum ML algorithm for our data set. We have explored  a variety of ML algorithms and compared their performances side by side. The procedure followed in this paper is novel because we have used one ML algorithm as base and have used other ML algorithms to train the error of the base ML algorithm. The entire algorithm as well as our best fit algorithm (which we  name as MIML model) is made available online for use of physics community for different nuclear physics and astrophysical applications.

As mentioned above, in this paper we have given a step by step outline on the way we have arrived at our final ML algorithm. However a couple of points need to be kept in mind. Firstly the nature of our problem is such that, for prediction, we would always be doing interpolation, or extrapolation very close to the domain of test data. More precisely, for prediction, we will mostly be doing interpolation in $Z$. Also for a fixed $Z$ we will do near extrapolation on
$N$ for neutron-rich elements, which may sometimes be difficult to produce in labs. That is reflected in the way we test and validate the ML models in the paper. Secondly, we have not obviously covered all aspects of ML algorithms employed in this paper. A more interested reader can refer to \cite{burkov2019hundred}.

However at the outset, we would like to clarify, that the ML model presented here is not in any way
comparable to a  physics model. A physics model of the nucleus is based on physical reasoning and there is a physical meaning to the small number of parameters present in the theory. Moreover a proper nuclear model should not only predict the BE but also other finite nuclear properties like nuclear charge radii, half lives and so on. Moreover a purely machine learning based model like ours will have many more parameters than a physics model and we will not be able to associate any physical meaning to those parameters. The goal of this work is not to compete with physics models but to explore ML algorithms and how accurately ML based models are able to make predictions even in the absence of inputs from physics models.

The paper is organized as follows. In Section \ref{sec2} we give a brief outline of the all machine learning algorithms employed. The results of this work is demonstrated in Section \ref{sec3} and we conclude in \ref{sec5}. We have three important appendices in our paper. Appendix \ref{algo} continues a technical description of our algorithm while Appendix \ref{programoverview} gives a detailed description on how to use our code to obtain the BE of any set of nuclei required by the user. Appendix \ref{datatable} gives the data for constructing the neutron star crust equation of state.

\section{Machine Learning (ML) Algorithms}
\label{sec2}
The ML can be extremely helpful to build statistical models on experimental data and then use them for predicting some property about newer data obtained by further experiments. Machine learning can be broadly divided into two parts:
\begin{itemize}
  \item Supervised Learning
  \item Unsupervised Learning 
\end{itemize} 
In supervised learning, the data that we have has two components, features and labels. The task we need to accomplish is to make a model which can predict the label of a set of new features, based on the available data of features and their corresponding labels. Labels are also known as the "target variable". The two main type of problems that can be solved using supervised learning are classification and regression problems. In classification problems, the target variable is a set of finite and discrete values called 'classes', and the task is to build a model which can assign the set of data points to one of the classes. In a regression problems, the target variable is a continuous valued variable, and the task is to build a model which can estimate the value of the target variable for a given set of features. In unsupervised learning, the data that we have is unlabeled, and the task is to build models that transform the data into some useful information depending on the problem we are trying to solve.

The problem we are trying to solve falls in the category of a supervised learning, regression problem. Our features are the atomic number (Z) and neutron number (N) of the nuclei and the target variable is the binding energy of the nuclei. So our task is to build a model using a labeled data set, that can estimate the value of binding energy of a nucleus given its atomic and mass number.

{To fit a machine learning model to a data set, we first divide the data set into three components - training set, validation set and testing set. The training set is a set of data points to which the model is fit. Since there is a chance that our choice of hyper-parameters was such that the model could over-fit the training set, we need a validation set which can give us an unbiased estimate of error by the model. By looking at the error on the validation set, hyper-parameter tuning is performed. If a change in a hyper-parameter results in a drop in validation error, it implies that the model has improved, because for the model, the validation set was completely new. After hyper-parameter tuning is complete, the final model is evaluated using the test set which was neither used for training, nor for parameter tuning. These results are considered as the final evaluation of the model.}

The major algorithms that can be employed to build regression models from labeled data are : 

{\bf Linear Regression (LR)-} Linear Regression tries to fit an equation of the form $y = a_0 + a_1x_1 + a_2x_2 + a_3x_3 + ...$ to the data where $y$ is the target variable and $x_i's$ are the features. The algorithm uses gradient descent on a cost function in the parameter space to find an optimal set of parameters $a_i's$. In our case, the equation will be of the form $BE = a_0 + a_1Z + a_2N$.

{\bf Decision Tree (DT)-} Decision tree works by dividing the data set into smaller parts which are similar in nature based on metrics like information entropy, variance, and impurity. It can fit non-linear functions because it basically works by dividing the entire feature space into cuboids and then making independent predictions in those cuboids.

{\bf Random Forest (RF)-} Random Forests are an ensemble of decision trees. They generally perform better than decision trees because they average out the errors made by individual decision trees in the ensemble\cite{10.1023/A:1010933404324}. 

{\bf Polynomial Regression (PR)} In polynomial regression, we try to fit a n-degree polynomial to the data. For that, we create all polynomial features of degree less than or equal to n (For example, features of degree less than or equal to 2 would be $1,Z,N,Z^2, N^2, ZN$) and then perform linear regression on that feature space.

{\bf Support Vector Machines (SVM)-} Support Vector Machines (SVM) make non-linear regression very easy because of the kernel trick. The SVM algorithm works by computing a similarity measure between two points in the feature space, we can define this similarity measure using a kernel which will effectively map all the points into a higher dimensional feature space, in which our data can be linear\cite{NIPS1996_1238}. The most popular kernel is the Gaussian (or radial basis function (rbf) ) kernel because it maps our feature space to an infinite dimensional space.
The rbf kernel has the following form: 
\begin{center}
    $k(x_i, x_j) = exp(- \gamma ||x_i - x_j||^2)$    
\end{center}

{\bf Error training on base ML algorithm-}
      We have designed a error training algorithm over ML algorithms as a base predictor, we expect that it may be possible to estimate the error by the base algorithm using another machine learning algorithm on top of it. This additional algorithm, which will be trained on the difference of the actual and predicted values (by base algorithm) of binding energy might capture some features of the data which the base algorithm failed to capture. This process may be repeated until the error becomes completely random and unpredictable by the ML algorithms. We used Random Forest for making error estimation algorithms because it can be applied to a wide range of data distributions. 
{It is to be noted that Random Forest are used for both classification and regression problems. \cite{10.1023/A:1010933404324} \cite{Genuer2008RandomFS} }.
Since in our case, the pattern of error will keep on changing as we iteratively subtract the error predictions from previous models, it becomes important
to use an algorithm that can fit to a wide range of data distributions. We used a validation set to keep track of the error after each iteration of error estimation and counted the number of iteration required (depending on the base algorithm) for the error to become completely random, then we use those number of iterations of error estimation on the test data.
\\
Error training on base ML algorithm can also be understood as an instance of \textit{Stacked Generalization}\cite{article} \cite{Breiman1996StackedR}. To quote Ref. \cite{article}, \textit{“Stacked generalization is a generic term referring to any scheme for feeding information from one set of generalizers to another before forming the final guess.”}. Our method perfectly fits this definition. According to the terminology used in Ref. \cite{article}, our base model is a “level-0” estimator, and the $n$ Random Forest models that we are using are “level-1”, “level-2”.... “level-n” estimators. For our base model (level-0 estimator), the output space is the values of $BE$, because it is trained on the AME2016 data set directly to predict the $BE$ from $Z$ and $N$, but for our level-1 estimator, the output space is the \textit{Errors of level-0 estimator}. Because our first RF model (level-1) is trained to predict the differences in level-0 predictions and actual AME2016 data. So, our level-1 estimator corrects for the errors made by the level-0 estimator. Similarly, the level-2 estimator corrects for the errors made by the level-0 estimator and level-1 estimator combined. Therefore, the level-2 estimator’s output space is the \textit{Error of level-0 and level-1 estimator combined}.  More Generally, for the level-x estimator, the output space is the \textit{Error of level-0, level-1 …. level-(x-1) estimators combined}.\\
An interesting feature of our procedure is that, we don’t fix the number of levels above level-0 (denoted by $n$)
beforehand. We treat this $n$ as a hyper-parameter and tune it for different level-0 estimators using a validation set because the optimal value of 
$n$ depends on the algorithm used at level-0.

\begin{figure}[htp]
        \centering
        \includegraphics[width=.5\textwidth,angle=0]{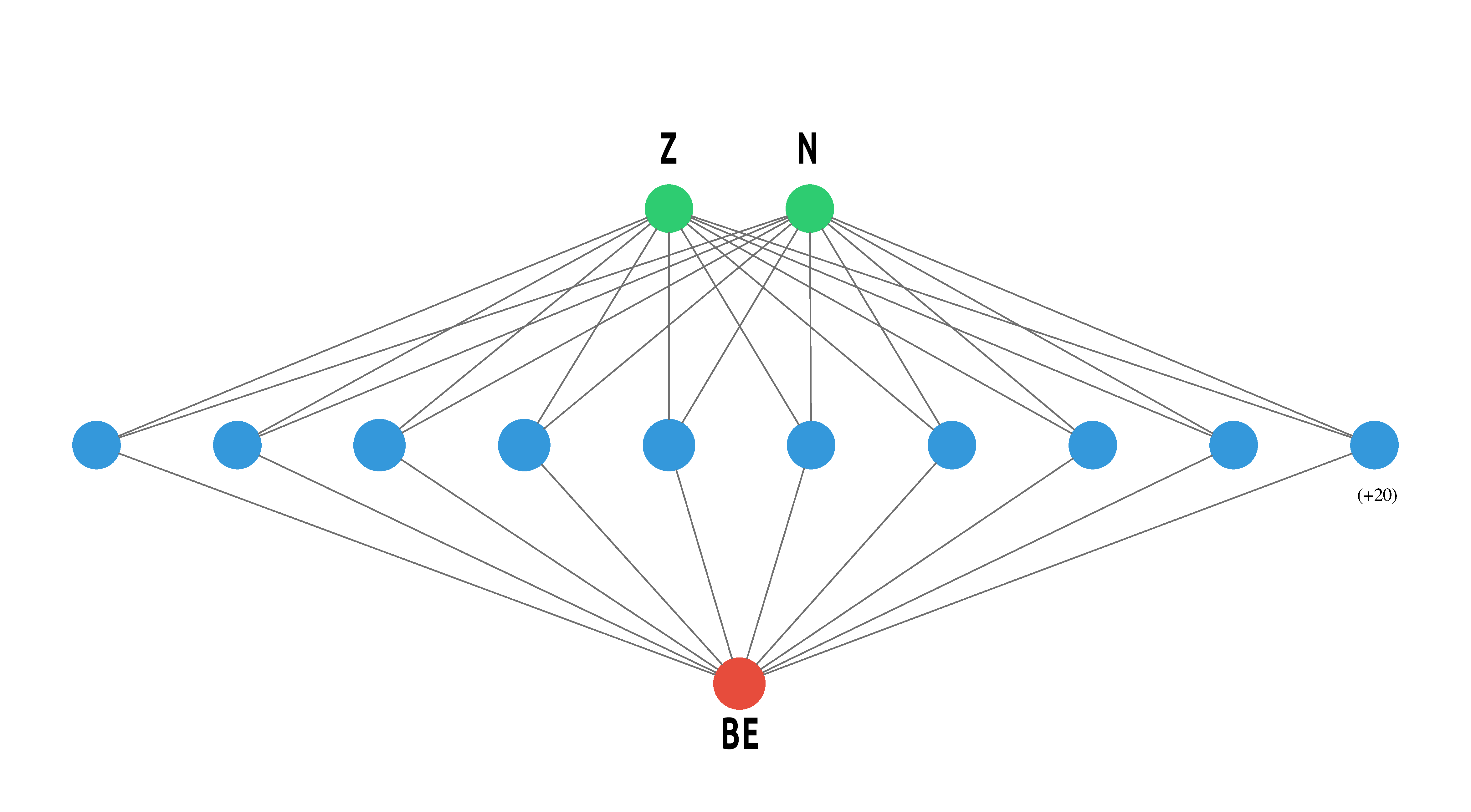}
           \caption{\label{schematic_nn} The schematic diagram of an artificial neural network (ANN). }
      \end{figure}

      {\bf Neural Network-} Neural networks are a class of machine learning algorithms that are loosely inspired by neurons in the human brain. The neural network is also a powerful tool to understand the complex dependency in the data. It is a mathematical function that maps a given input to the desired output.     
      A neural network consists of hierarchical layers made of neurons (the basic unit). We will consider neurons with a vector of $I$ input signals $\mathbf{x} = \left\{ x_i \right\}_{i=1}^I$ (in our case, $I=2$ and $x_1=Z$ and $x_2=N$) and an output signal $y(a)$ (in our case $y=BE$), which is a (often non-linear) function of the {\it activation}
      $a = \sum_i w_i x_i,$ where $\mathbf{w} = \left\{ w_i \right\}_{i=1}^I$ are the weights of the neuron. The sum runs from either 1 to $I$, or from 0 to $I$ if there is also a bias ($b \equiv w_0$). The architecture of our neural network is shown in Figure \ref{schematic_nn}. It contains one hidden layer having 30 nodes, each activated by a ReLU activation function.
      Any {\it activation} function can be chosen depending upon the problem at hand. Some commonly used activation functions are: {\it elu}, {\it softmax}, {\it selu}, {\it softplus}, {\it softsign}, {\it relu}, {\it tanh}
      , {\it sigmoid}, {\it linear} and {\it exponential}. Training the neural network involves finding the weights and biases by minimizing a loss function given the training data. 
      The choice of the loss function depends on the type of data and the desired prediction. Depending upon the non linearity of the problem one can use many layers of neurons each with different number of nodes. The choice of the number of layers and nodes is the art of optimization. 
         
     Since we are interested in minimizing the mean squared error, in this work we have used the loss function given below,
      \begin{equation}
       \mathcal{L}(\mathbf{w},b)=(1/N)\sum_i(\hat{y_i}(\mathbf{w},b) -y_i)^2
      \end{equation}
      where $y_i$ is the experimental data and $\hat{y_i}(\mathbf{w},b)$ is the prediction of neural network. 
      
      For this work we have implemented ML and Artificial Neural Network(ANN) algorithms by using 
      Python packages TensorFlow2.0\cite{tensorflow2015-whitepaper}, Keras\cite{chollet2015keras}, and Scikit-Learn\cite{scikit-learn}.

\section{Results} 
\label{sec3}
{We construct the NS outer crust equation of state based only on Machine Learning algorithms on experimental data of the 3435 nuclei given in AME2016 \cite{Wang2017}. However, as we already stated in the introduction,  the aim is not to compete with physics models but to explore ML algorithms and determine how accurately ML based models are able to make predictions even in the absence of inputs from physics models. To determine the composition of the NS outer crust the knowledge of nucleus binding energy for different nucleus are essential as the Gibbs free energy or chemical potential needs to minimize at a certain pressure with respect to the atomic mass number A and the atomic charge Z. We will first demonstrate different types of ML algorithm and their limitation to predict atomic BE. Next we will construct the NS outer crust EoS based on our best ML algorithm. }

\subsection{Nuclear Binding Energy (BE) based on Machine Learning}
We do not assume any nuclear physics models in our work. We predict nuclear BE based only on Machine Learning algorithms. Our data set is AME2016 \cite{Wang2017}, where we only look at $Z$, $N$ and $BE$ data of all nuclei. We break our data set randomly into 60\% training data 20\% validation and 20\% testing set. In the first part we chose the base algorithm of ML which gives the lowest $\sigma_{\rm rms}$. In the second part we train the error obtained from the base algorithm to further reduce the $\sigma_{\rm rms}$ on the test set. 
      
\begin{figure*}[htp]
\centering
\begin{tabular}{cc}
\includegraphics[width=.45\textwidth,angle=0]{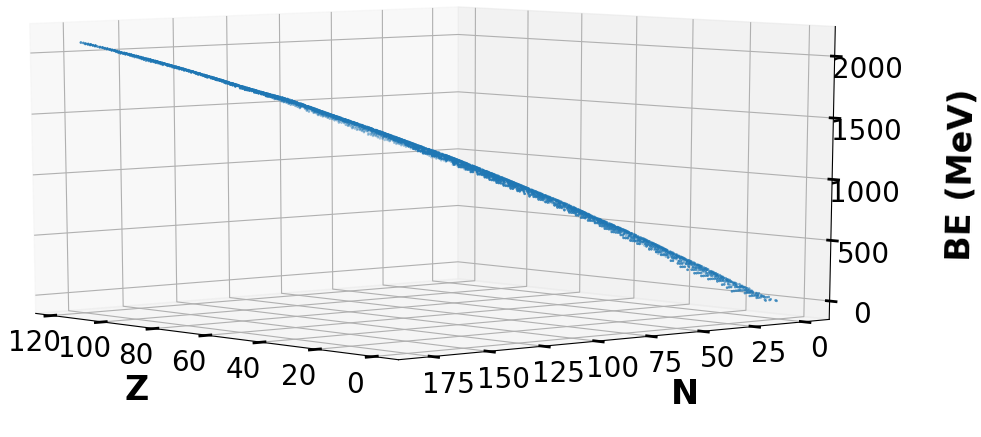}
&
\includegraphics[width=.45\textwidth,angle=0]{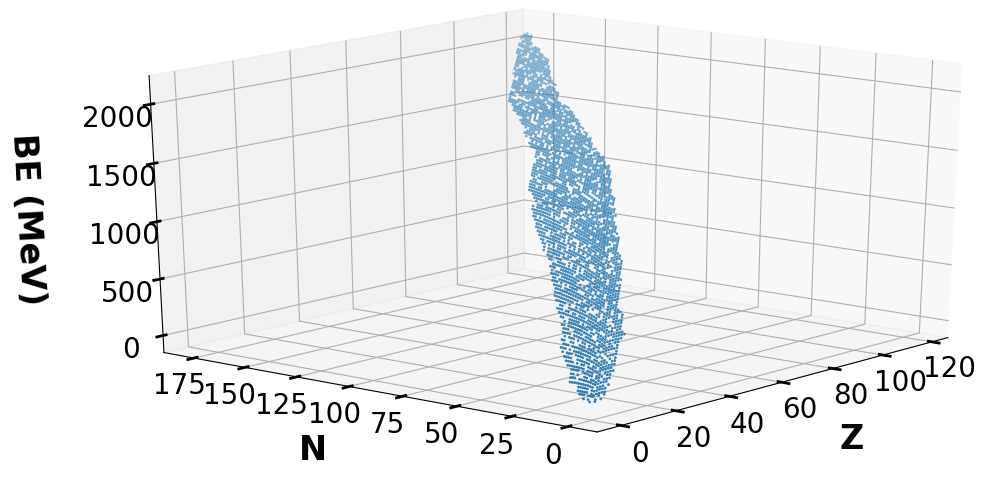}
\end{tabular}
\caption{\label{raw_data} (left) The 3D variation of binding energy (BE), (right) 3D view of the same from different angles with respect to $Z$ and $N$ of AME2016  data \cite{Wang2017}.}
\end{figure*}

The first step in applying machine learning to any data, is to visualize the distribution of the data. The plot of the  binding energy vs. (Z,N) is given in Figure \ref{raw_data}.
      From Figure  \ref{raw_data} (right) it seems that the data is fairly linear and can be fit well by a simple linear regression, however, upon closer inspection, 
it can be seen that the plot is actually slightly curved, and we need nonlinear models to fit the data. The curvature in the plot can be observed from a particular orientation 
of the axes as shown Figure  \ref{raw_data} (left). It is evident from the figure that non-linear models will fit the data better than linear regression, 
but the error in linear regression can be used as a benchmark for other non-linear models.  
      
\begin{figure*}[htp]
\centering
\begin{tabular}{ccc}
\includegraphics[width=.3\textwidth,angle=0]{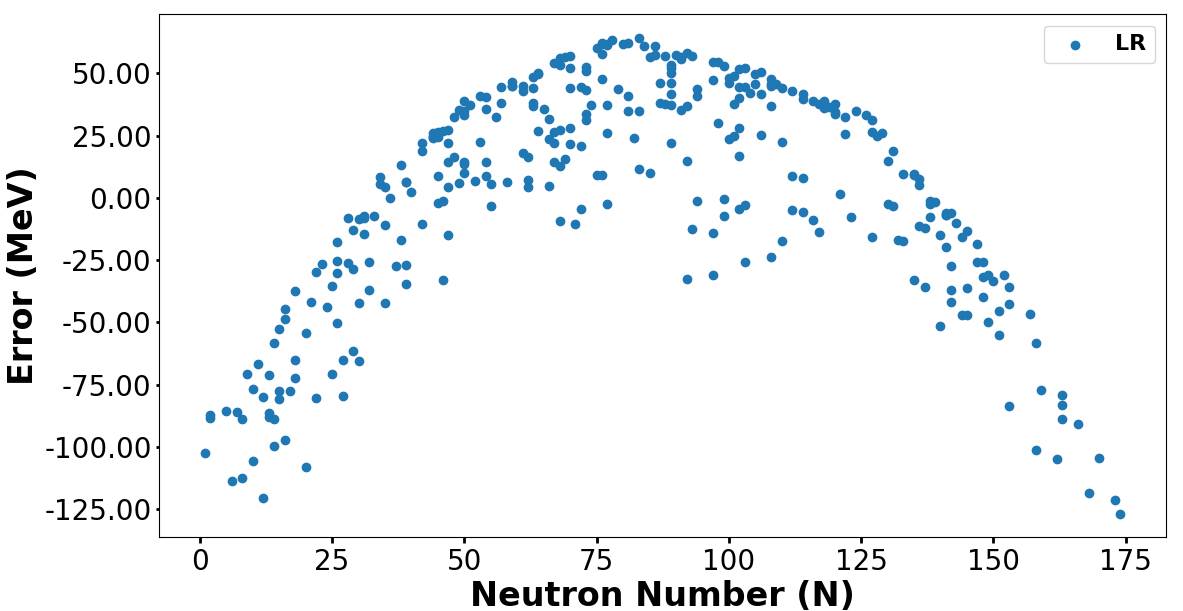}
&
\includegraphics[width=.3\textwidth,angle=0]{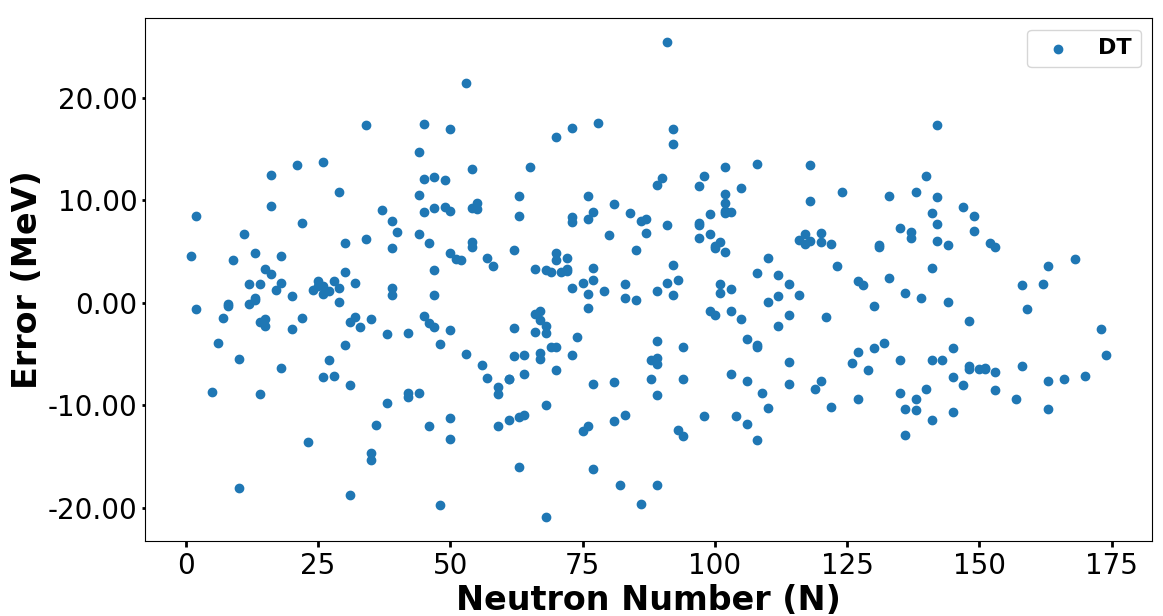}
&
\includegraphics[width=.3\textwidth,angle=0]{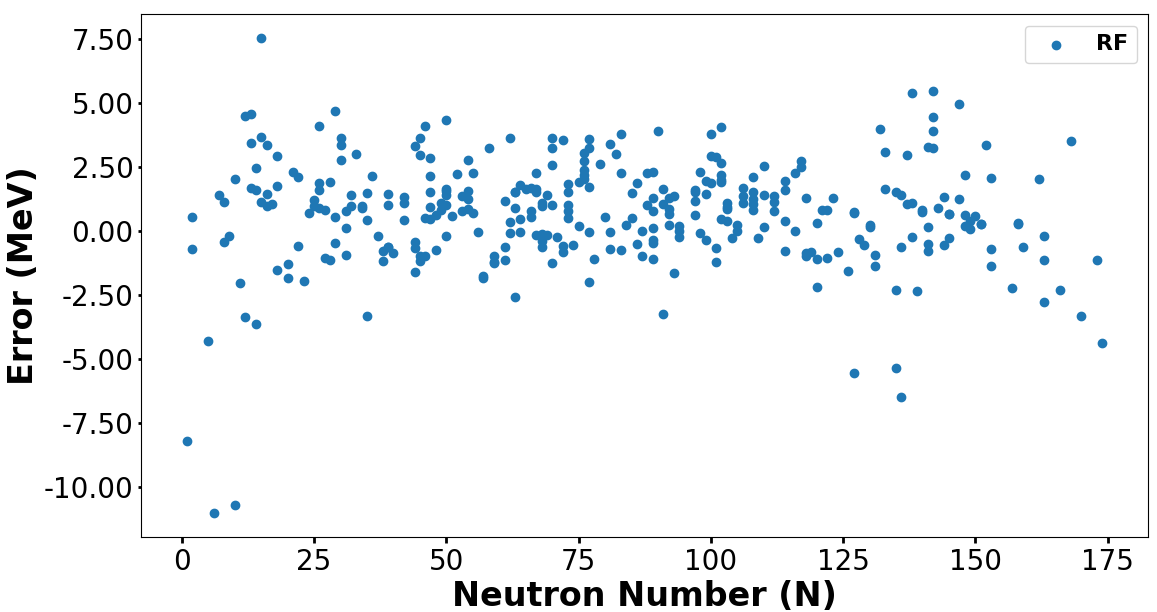}
\end{tabular}
\begin{tabular}{cc}
\includegraphics[width=.3\textwidth,angle=0]{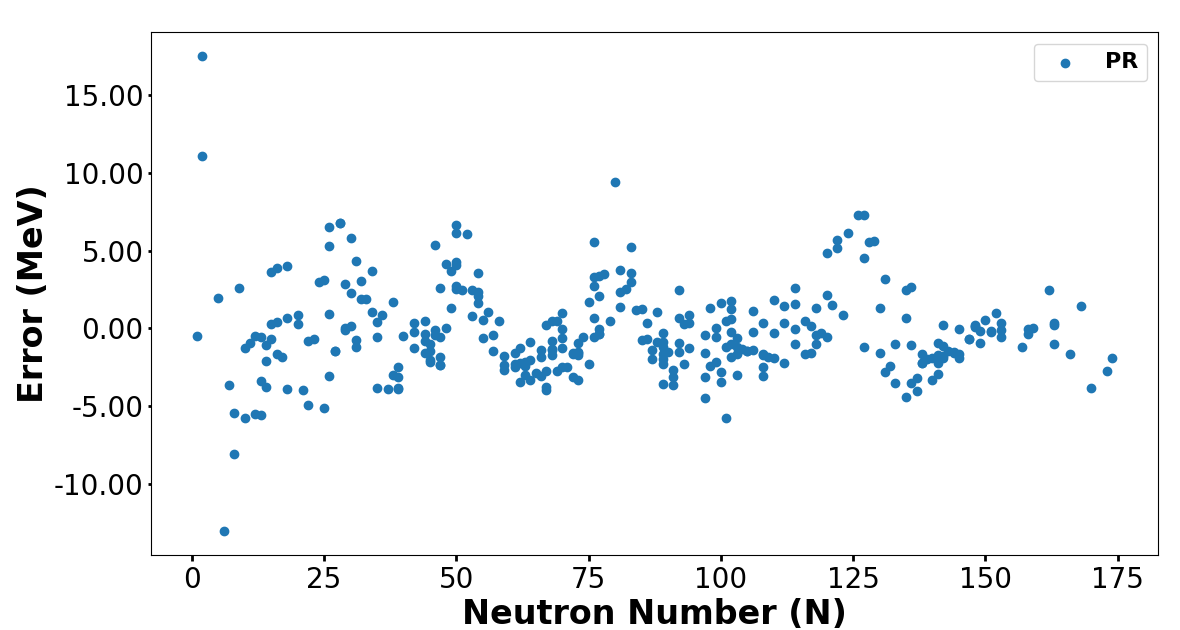}
&
\includegraphics[width=.3\textwidth,angle=0]{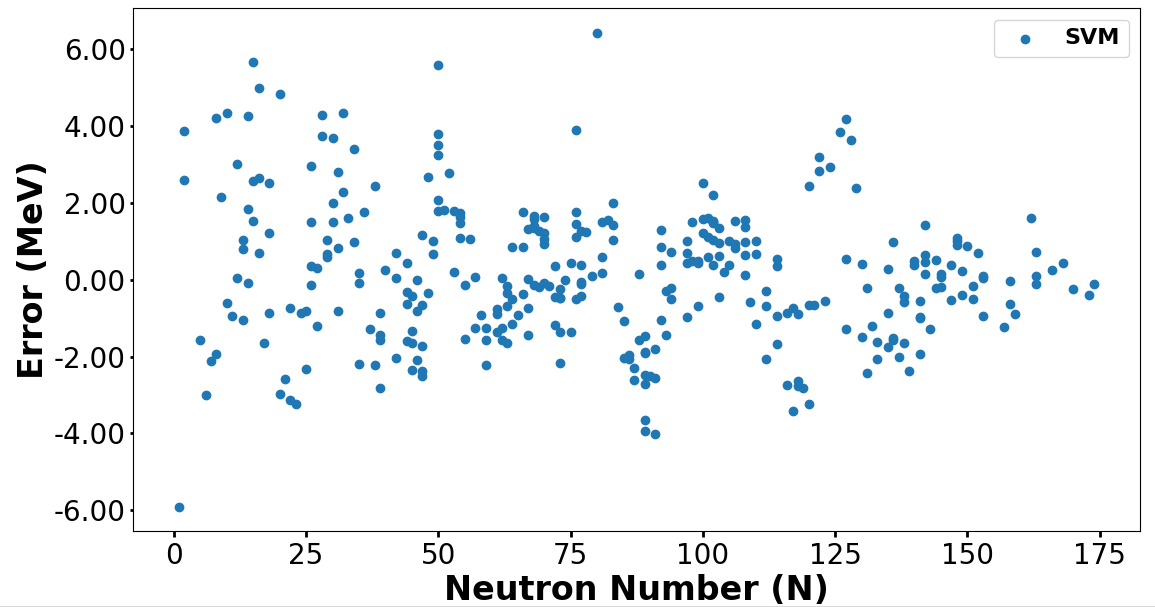} 
\end{tabular}
\caption{\label{base_err} Error in individual data points corresponding to neutron number (N) calculated for (top left) Linear Regression, (top middle) Decision Tree, (top right) Random Forest, (bottom left) Polynomial Regression and (bottom right) SVM base algorithms of AME2016 data set.}
\end{figure*}
The RMS error ($\sigma_{\rm rms}$) on the test set on using LR comes out to be $\sim 45.78$ MeV. The linear regression is given by the equation
${\rm BE} = 8.3~Z + 6.7~N$. This error will now serve as a benchmark for us, because any non-linear model should perform better than this on our data. If that is not the case, it implies that we need to tune the hyper-parameters of that model. We also plot the prediction error against the neutron number (N) of the nuclei in Figure \ref{base_err} (top left). 
In DT, the $\sigma_{\rm rms}$ on the test set reduces and is $\sim 8.22$ MeV which is very less as compared to linear regression. 
It has also captured the non-linearity. The graph of error vs N appears to be very random and does not show any pattern upon inspection (see Figure \ref{base_err} (top middle)).
The RF fits the data better than decision trees, it gives an $\sigma_{\rm rms}$ of $\sim 2.18$ MeV on the test set. The graph of error vs $N$ for random forest is shown in Figure \ref{base_err} (top right). 
It is more concentrated as compared to decision tree graph which shows that the prediction has improved but still we can't clearly see a pattern in the error. In case of PR, the degree of the polynomial is a hyper-parameter which needs to be tuned. We have tried out various degrees and tested it on a validation set and saw the trend in the error and then have chosen the degree with the least error. The error on validation set becomes minimum for the degree 6, therefore we have tried to fit a 6 degree polynomial to the training set. The $\sigma_{\rm rms}$ on test set for polynomial regression is $\sim 2.58$ MeV. The graph of error vs $N$,(Figure \ref{base_err} (bottom left)) shows a pattern. It peaks at certain values of N and remains close to zero for others. By using Gaussian kernel in SVM (parameters: $C = 5\times 10^5, \gamma = 5\times10^{-4}$) the error by SVM on test data was $\sim 1.81$ MeV, which is lowest among all models which were tried out. The error vs $N$ plot is shown in Figure \ref{base_err} (bottom right). It shows similar pattern similar to what was obtained for polynomial regression.

Since we observe that a specific pattern arises on plotting error against neutron number $N$ (see Figure \ref{base_err}), it is justified 
to train these error by another ML algorithm to reduce the $\sigma_{\rm rms}$.
We train our Base Model on the training set and obtained the Root-Mean-Squared-Error ($\sigma_{\rm rms}$) on the testing set, which we denote as $\sigma_{\rm rms}^i$. By using a non-linear error estimating algorithm on top of base algorithms, we can further capture the non-linearity of the data. The algorithm which is used in error training for our data set is Random Forest, as justified in {Section \ref{sec2}}. The number of Random Forest models, $n$ is considered on top of the base model until the $\sigma_{\rm rms}$ saturates. The final $\sigma_{\rm rms}$ is calculated on the testing set and is denoted by $\sigma_{\rm rms}^f$.
\begin{figure*}[htp]
\centering
\begin{tabular}{ccc}
\includegraphics[width=.3\textwidth,angle=0]{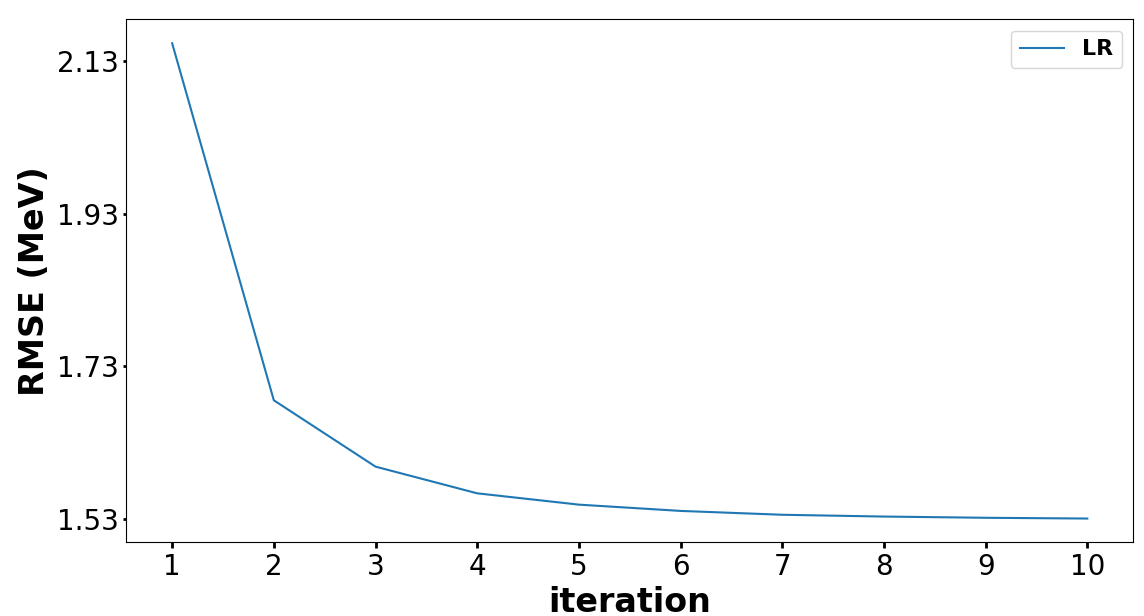}
&
\includegraphics[width=.3\textwidth,angle=0]{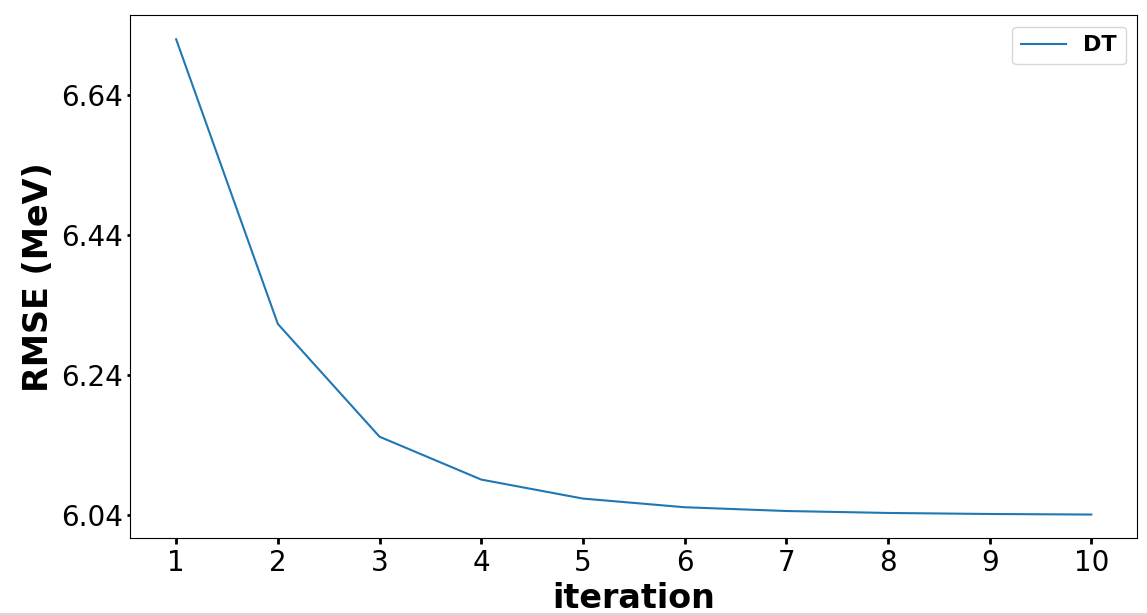}
&
\includegraphics[width=.3\textwidth,angle=0]{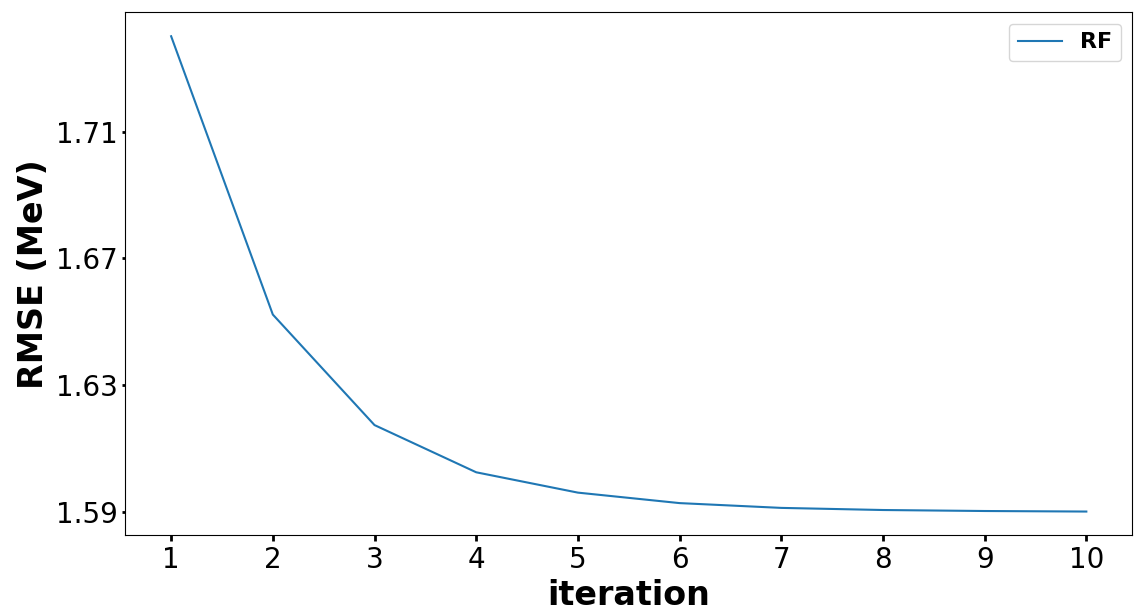}
\end{tabular}
\begin{tabular}{cc}
\includegraphics[width=.3\textwidth,angle=0]{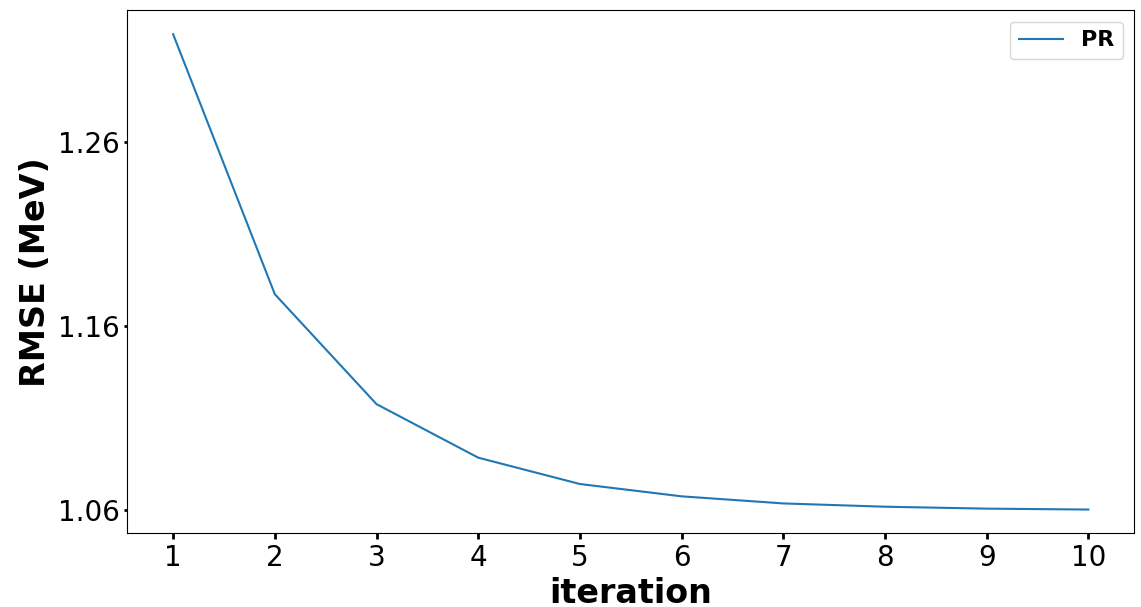}
&
\includegraphics[width=.3\textwidth,angle=0]{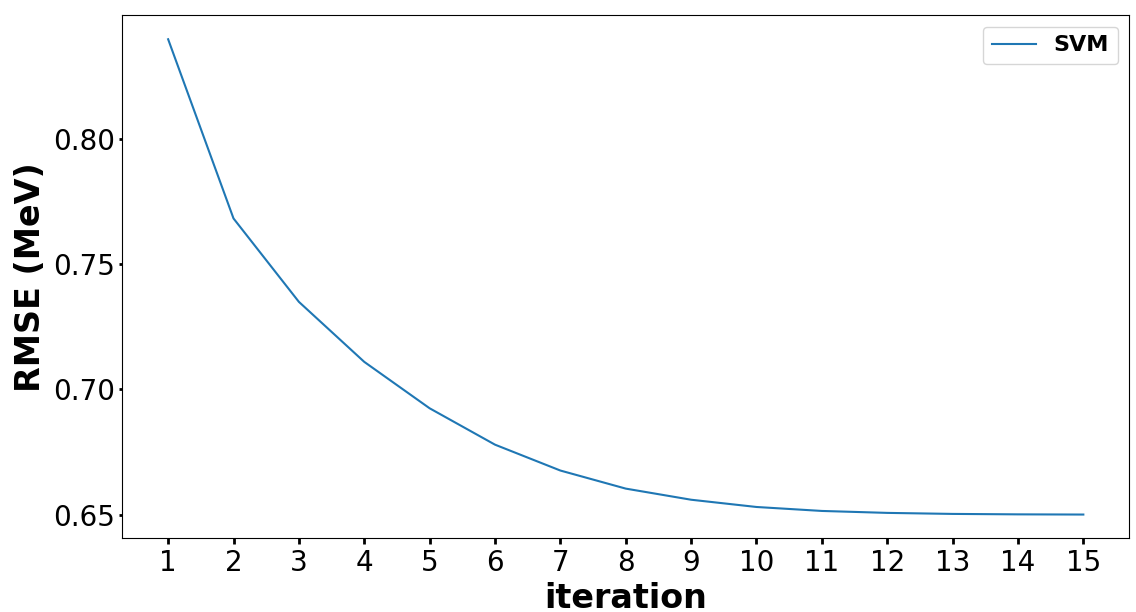} 
\end{tabular}
\caption{\label{err_train} The plot for the $\sigma_{\rm rms}$ of combined model on validation set versus number of Random Forest
iterations on top of base models (top left) Linear Regression, (top middle) Decision Tree, (top right) Random Forest, (bottom left) Polynomial Regression and (bottom right) SVM base algorithms of AME2016 data set.}
\end{figure*}      
In Figure \ref{err_train} we plot the $\sigma_{\rm rms}$ vs the number of iterations in error training for all base ML algorithms.

\begin{figure*}[htp]
\centering
\begin{tabular}{cc}
\includegraphics[width=.4\textwidth,angle=0]{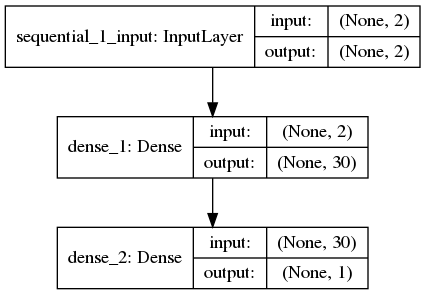}
& \includegraphics[width=.5\textwidth,angle=0]{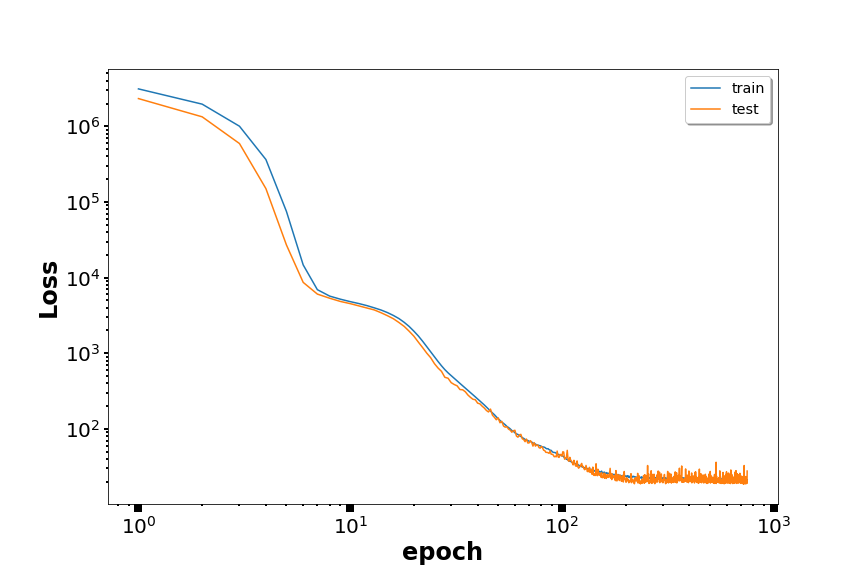}
\end{tabular}
\caption{\label{ann} (left) A schematic diagram of our ANN algorithm (right) and the plot for the loss function verses number of epoch of the ANN model. }
\end{figure*}

Apart from the above-mentioned Machine Learning algorithms, we also use artificial neural networks(ANN) to make predictions of
the BE. We again use AME2016 data set and followed the same treatment to train, validate  and test the neural network. 
It is to be noted that, currently, there are no rules on selecting the proper number of layers and nodes but is more of
an art in network training. For our data set, we managed to obtain a $\sigma_{\rm rms} \sim 5 MeV$ on test sets for the
best optimized ANN which is higher than our best results using Machine Learning algorithms. Instead of using an ANN
alone we have improved the network by having a Linear Regression base model and training the ANN on that. The lowest 
$\sigma_{\rm rms}$ we have got via this approach is $\sigma_{\rm rms}=2.30$ MeV. The network architecture is plotted
in Figure \ref{ann} (left). The Loss function chosen is 'mean squared error' and is plotted for both test and train data as 
a function of epoch in Figure \ref{ann} (right). 
This shows the network loss function is properly minimized. 
As mentioned above, in the  prediction of  BE of a nuclei directly via learning algorithm, the Machine Learning
algorithms gives better performance rather than Neural network. 

The results of all the base ML algorithms along with subsequent error training by Random Forest is summarized in 
Table \ref{tab_out}. {The reported values of $\sigma^f_{rms}$ for various models vary within $\pm$ 0.1MeV if we choose different sets
for training and testing for each of the above-mentioned ML algorithms.} As can be seen from Table \ref{tab_out}, the best results are obtained from base model 
SVM and subsequent error training by random forest up to $n=10$. This model will be referred to as the MIML model in the rest of the paper.  
We have avoided over-fitting the models by having a validation set (which was not a part of the training set) and keeping track of the error on validation set. If the error on the training 
set becomes very low and the error on the validation set begins to diverge, it shows that the model has been over-fitted.  We have considered only those many Random Forest 
models on top of the base model, for which the validation error did not diverge. However, there is still a possibility for the model to over-fit in the very first Random Forest iteration 
itself, so to avoid that, we have fixed the maximum depth of the estimators in Random Forest to be 30. This will prevent over-fitting in the first iteration itself, and then we control the 
number of Random Forest models to handle over-fitting for the combined model.
{As seen in Figure \ref{err_train}, the validation error is converging after the number of RF models
that we consider in the combined model, which clearly indicates that the combined model does not over-fit}.
\begin{table}[htp]
\caption{\label{tab_out} The $\sigma_{\rm rms}$ error of base ML models ($\sigma_{\rm rms}^i$) as well as error trained model ($\sigma_{\rm rms}^f$) is outlined.
$n$ Random forest models on top of the base model are considered for error trained model. All the errors are for the test set
only} 
\centering
\setlength{\tabcolsep}{14pt}
\renewcommand{\arraystretch}{1.1}
\begin{tabular}{cccc}
\hline \hline 
Base Model            & $\sigma_{\rm rms}^i$ & $n$ & $\sigma_{\rm rms}^f$ \\
                      &  (MeV)           &     &  (MeV)      \\ \hline 
LR    & 45.78 & 3                              & 1.65       \\
DT    & 8.22  & 6                              & 6.32       \\
RF    & 2.18  & 4                              & 1.56       \\
PR    & 2.58  & 4                              & 1.18       \\
SVM   & 1.81  & 10                             & 0.58      \\ \hline \hline 
\end{tabular}
\end{table}

\begin{figure}[htp]
\centering
\includegraphics[width=.8\textwidth,angle=0]{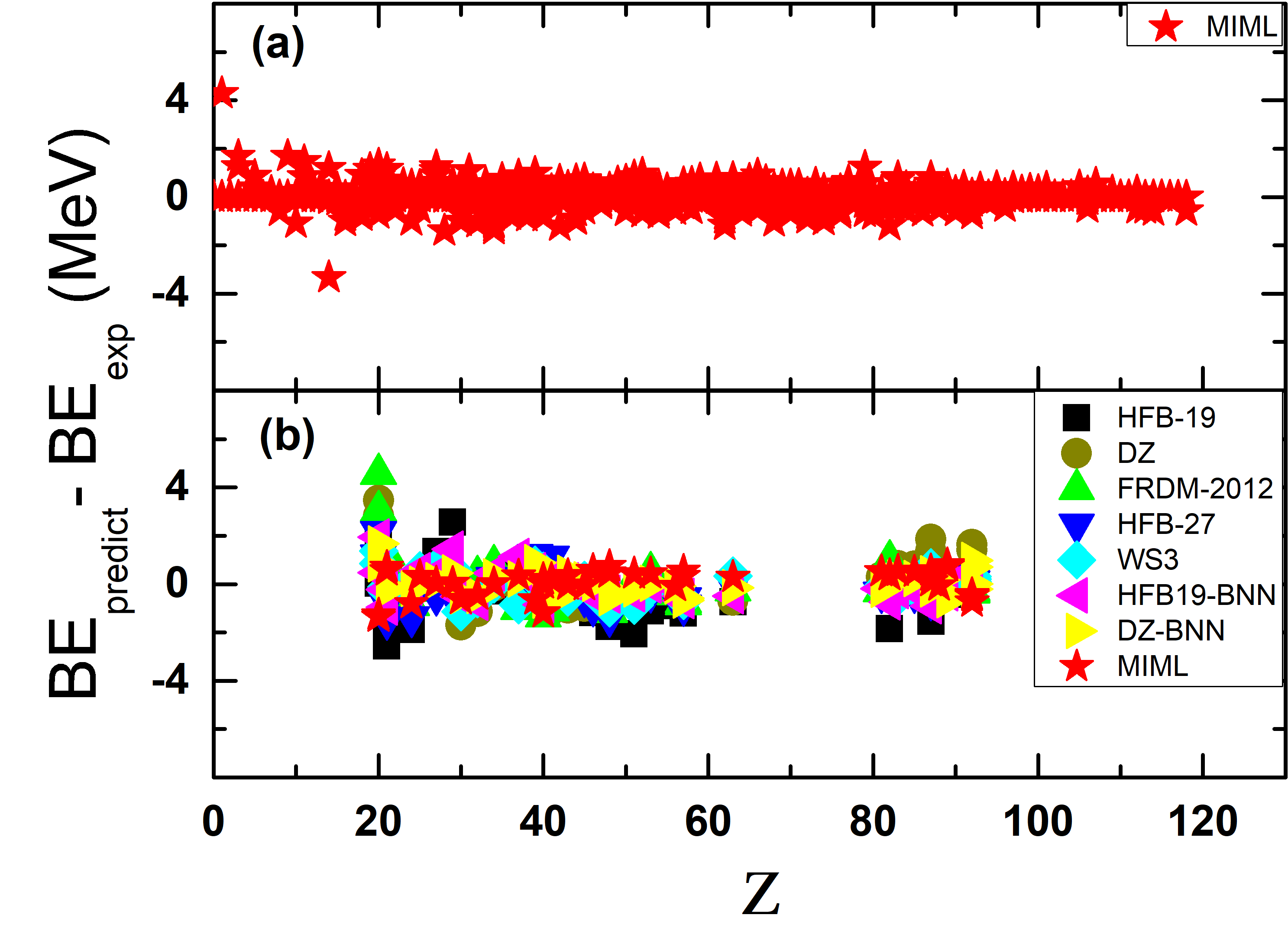}
\caption{\label{comp} (a) The difference of binding energy (BE) between MIML predicted algorithm and AME2016 data set verses proton number $Z$, (b) the difference of BE is compared with some well known nuclear models 
for newly compiled mass values for 46 additional nuclei in the $^{40}$Ca -  $^{240}$U region in AME2016 as compared to AME2012. It is to be noted that 
	to check the robustness of our algorithm those 46 nuclei was not part of the training set in this case.}
\end{figure}

In Figure \ref{comp}(top) we plot the difference between predicted BE using the MIML algorithm (the best stacked machine learning model in our trained model) and AME2016 BE data verses proton number $Z$.
As mentioned above, Our MIML algorithm is independent of any physics model. The only physics input going into this is
that a nucleus is characterized by $Z$ and $N$. 
Most mean field theories of nuclear models predicts BE with high errors for very light nuclei and also for heavy
nuclei which have a large proton neutron number asymmetry. As a result any
machine learning algorithm used on top of a physics based model will inherit the drawbacks of the model. Our algorithm,
being model independent, does not suffer from such weakness and we train and test on the entire data set of AME2016. 
As can be seen from Figure \ref{comp}(top) our predictions
match the actual BE values for all values of $Z$, including those for light, heavy as well as magic nuclei.
The $\sigma_{\rm rms}$ for light nuclei below $Z<20$ is 505 KeV. For nuclei between  $20\leq Z \leq 92$ the $\sigma_{\rm
rms}$ is 248 KeV  while for heavy nuclei $Z>92$ it is 127 KeV \footnote{ Note that these values are for the entire data set, i.e. it includes training, validation and test sets. Hence the error here is lower than the one reported in Table \ref{tab_out} which was for the test set only}.
{We have also calculated the $\sigma_{\rm rms}$ in the MIML model on different branches of nuclei in Table \ref{tab_rms}.}
\begin{table*}[]
\centering
\caption{\label{tab_rms} The $\sigma_{\rm rms}$ for nuclear BE obtained in MIML model for different branches of nuclei present in AME2016. The \textit{odd-odd} indicates the
nuclei with odd $Z$ and odd $N$. The similar nomenclature is for \textit{even-even}. The \textit{odd-even} is for those nuclei either with odd $Z$ and even $N$ or vice versa. The \textit{magic nuclei} are those nuclei which have $Z$ or $N$ as $2,8,20,28,50,82,126$.}
\setlength{\tabcolsep}{16pt}
\renewcommand{\arraystretch}{1.2}
\begin{tabular}{ccccc}
\hline \hline
                   & \textit{odd-odd} & \textit{odd-even} & \textit{even-even} & \textit{magic nuclei} \\
                   & (MeV)     &  (MeV)     &  (MeV)      & (MeV) \\ \hline
$\sigma_{\rm rms}$ & 0.0149    & 0.02       & 0.027       & 0.024 \\ \hline \hline
\end{tabular}
\end{table*}

In Figure \ref{comp}(bottom), we  compare our results with some well known theoretical models as well as theoretical base model with artificial neural network error training models.
To have a comparison on the same footing we choose the 46 nuclei in the $^{40}$Ca
-  $^{240}$U region which were not in AME2012 \cite{Wang2012} but was newly added in AME2016 (as was done in \cite{Utama2018}). For the purpose
of this comparison we do not include these 46 nuclei in the training set. 
Figure \ref{comp}(bottom) shows the comparison of ${\rm BE}_{\rm predict} - {\rm BE}_{\rm AME2016}$ of MIML algorithm with HFB-19 \cite{Goriely2010}, Duflo-Zuker \cite{Duflo1995}, FRDM-2012 \cite{Moller2012}, HFB-27 \cite{Goriely2013}, and WS3 \cite{Liu2011} as well as Bayesian Neural Network (BNN) improved Duflo-Zuker 
and HFB-19 models \cite{Utama2018} for these 46 nuclei. In Table \ref{tab_compare} we compare the overall $\sigma_{\rm rms}$ of the MIML model with that of the above mentioned models for these 46 nuclei.
%
\begin{table*}[htp]
\caption{\label{tab_compare} $\sigma_{\rm rms}$ of the predictions of various models for the 46 nuclei in the $^{40}$Ca - $^{240}$U region that appear in the latest
 AME2016 \cite{Wang2017} compilation but not in AME2012 \cite{Wang2012}.}
\setlength{\tabcolsep}{5pt}
\renewcommand{\arraystretch}{1.2}
\begin{tabular}{ccccccccc}
\hline \hline 
Model & HFB-19 & DZ    & FRDM-2012 & HFB-27 & WS3   & HFB19-BNN & DZ-BNN & MIML \\ \hline 
$\sigma_{\rm rms}$  & 1.093  & 1.018 & 0.997     & 0.723  & 0.513 & 0.587     & 0.479  & 0.501 \\
\hline
\hline
\end{tabular}
\end{table*}
From \ref{tab_compare} it is clear that the overall $\sigma_{\rm rms}$ on our MIML model is less than all the theoretical models and is very close to the Duflo-Zuker BNN improved model. In the next section we will use this MIML model to predict nuclear BE for NS outer crust EoS calculation.

\subsection{Neutron Star outer crust EoS}
The NS outer crust densities range from $10^{4} \mathrm{~gm} / \mathrm{cm}^{3}$ to a neutron drip density of about $4 \times 10^{11} \mathrm{~gm} / \mathrm{cm}^{3}$ and the composition is determined by minimizing the the Gibbs free energy or chemical potential at a certain pressure with respect to the atomic mass number A and the atomic charge Z. In the outer crust (i.e., before neutron drip) the energy per nucleons consists of three different contributions:
nuclear, electronic, and lattice.
\begin{equation}
\varepsilon(A, Z ; n_B)=\varepsilon_{n}+\varepsilon_{e}+\varepsilon_{\ell}
\end{equation}
where the baryon density is denoted by $n_B \equiv A / V$. The nuclear contribution to the total 
energy per nucleons is independent of the density. It is given by
\begin{equation}
\varepsilon_{n} \equiv \frac{M(Z,N)}{A}
\end{equation}
with
$$
M(Z,N)=N m_{n}+Z m_{p}-BE(Z,N)
$$
Here $M(N, Z)$ is the nuclear mass, $BE(N, Z)$ is the corresponding binding energy, and $m_{n}$ and $m_{p}$ are neutron and proton masses, 
respectively. This is the quantity we need to obtain from the nuclear mass models or from experimental data.

The electronic contribution reads $\varepsilon_{e} =\mathcal{E}_{\mathrm{el}} V,$ where the energy density $\mathcal{E}_{\mathrm{el}}$ of the electrons can be considered as that of a degenerate relativistic free Fermi gas and is given by
\begin{eqnarray}
	\mathcal{E}_{\mathrm{el}} \equiv & \frac{k_{\mathrm{F} \mathrm{e}}}{8 \pi^{2}}\left(2 k_{\mathrm{Fe}}^{2}+m_{\mathrm{e}}^{2}\right) \sqrt{k_{\mathrm{Fe}}^{2}+m_{\mathrm{e}}^{2}} \nonumber \\
&-\frac{m_{\mathrm{e}}^{4}}{8 \pi^{2}} \ln \left[\left(k_{\mathrm{Fe}}+\sqrt{k_{\mathrm{Fe}}^{2}+m_{\mathrm{e}}^{2}}\right) / m_{\mathrm{e}}\right]
\end{eqnarray}
with $k_{\mathrm{Fe}}=\left(3 \pi^{2} n_{\mathrm{e}}\right)^{1 / 3}$ being the Fermi momentum of the electrons, 
$n_{\mathrm{e}}=(Z / A) n_{\mathrm{b}}$ the electron number density, and $m_{\mathrm{e}}$ the electron rest mass. 
The lattice energy can be written as
\begin{eqnarray}
\varepsilon_l \equiv -C_{1} \frac{Z^{2}}{A^{1 / 3}} k_{\mathrm{Fb}} 
\end{eqnarray}
where $k_{\mathrm{Fb}}=\left(3 \pi^{2} n_{\mathrm{b}}\right)^{1 / 3}=(A / Z)^{1 / 3} k_{\mathrm{Fe}}$ is the average Fermi momentum 
and $C_{1}=3.40665 \times 10^{-3}$ for bcc lattices \cite{RocaMaza2008}.

The basic assumption in the calculation is that thermal, hydrostatic, and chemical equilibrium is reached in each layer of the crust. As no pressure is exerted by the nuclei, only the electronic and lattice terms contribute to the pressure in the outer crust. Therefore, we have
\begin{eqnarray}
P=-\left(\frac{\partial \varepsilon}{\partial V}\right)_{T, A, Z}=\mu_{\mathrm{e}} n_{\mathrm{e}}-\mathcal{E}_{\mathrm{el}}-\frac{n_{\mathrm{b}}}{3} C_{1} \frac{Z^{2}}{A^{4 / 3}} k_{\mathrm{Fb}}
\end{eqnarray}
where $\mu_{\mathrm{e}}=\sqrt{k_{\mathrm{Fe}}^{2}+m_{\mathrm{e}}^{2}}$ is the Fermi energy of the electrons including their rest mass. 
We have also used  $n_B \equiv A / V$ to obtain this. 
Now we need to find the nucleus that, at a certain pressure, minimizes the Gibbs free energy per particle, 
or chemical potential, $\mu=G / A=(E-T S) / A+P / n_B$. Since the Fermi energy of the electrons is much larger 
than the temperature of the star, to a very good approximation, we can take the temperature to be zero. The quantity to be minimized is given by
\begin{eqnarray}
	\mu(A, Z, P) &=&\frac{\varepsilon_n}{A} +\frac{P}{n_B} \nonumber \\
&=&\frac{M(N,Z)}{A}+\frac{Z}{A} \mu_{\mathrm{e}}-\frac{4}{3} C_{1} \frac{Z^{2}}{A^{4 / 3}} k_{\mathrm{Fb}}
\end{eqnarray}
The nuclear mass $M(N,Z)$ is the only unknown quantity in the above equations. Here we have calculated $M(N,Z)$ from the predicted $BE(Z,N)$ of MIML model.

\begin{figure}[htp]\centering
\begin{tabular}{c}
\includegraphics[width=0.95\textwidth,angle=0]{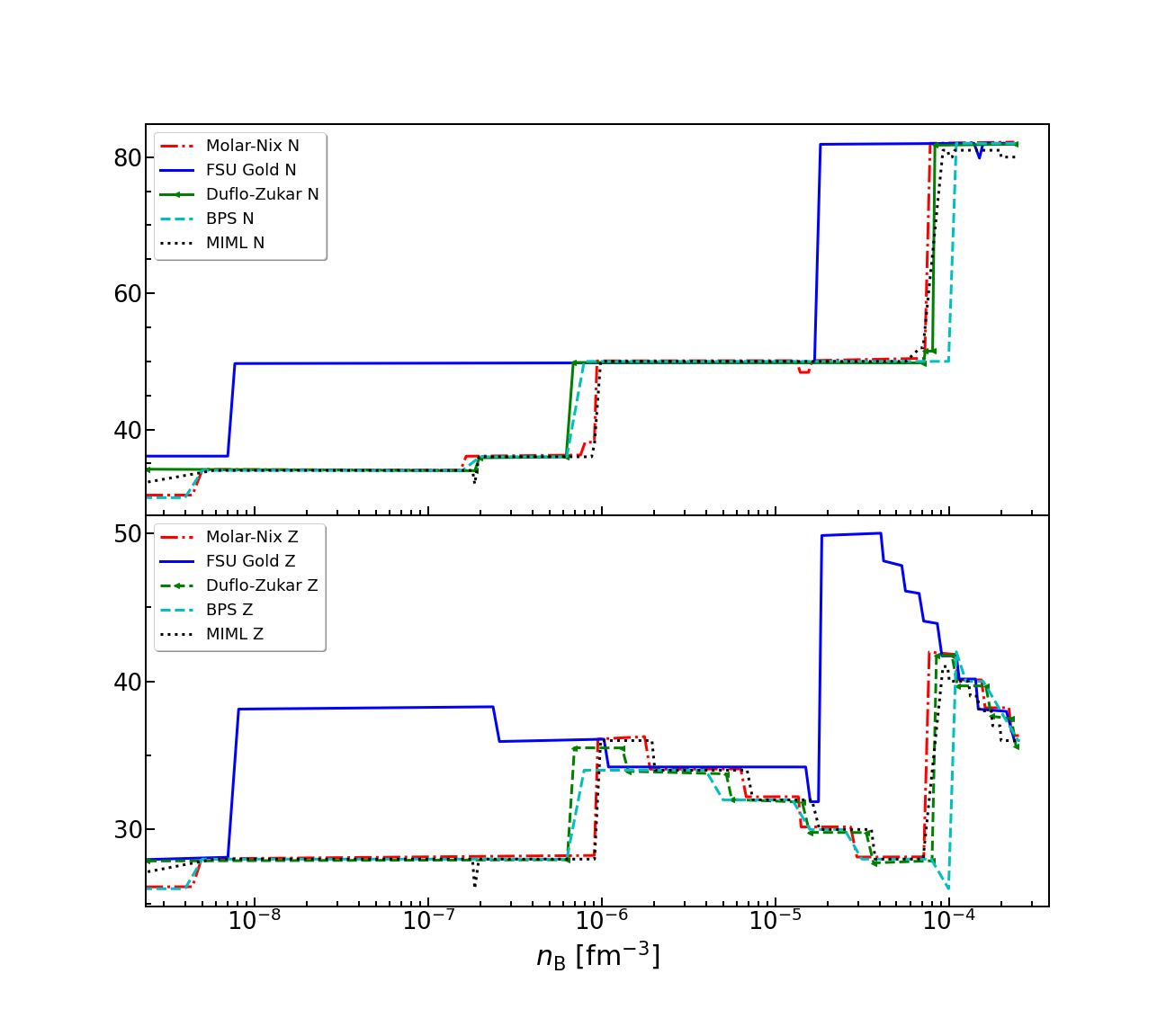} 
\end{tabular}
\caption{The neutron star outer crust atomic composition of different nuclei's
(upper panel) neutron number and (lower panel) proton number verses number density 
as predicted by using the mass formulae of Moller-Nix,  FSUGold, Duflo-Zuker, BPS and 
our MIML model.} 
\label{fig:crustcompfig}
\end{figure}

\begin{figure}[htp]\centering
\begin{tabular}{c}
\includegraphics[width=.80\textwidth,angle=0]{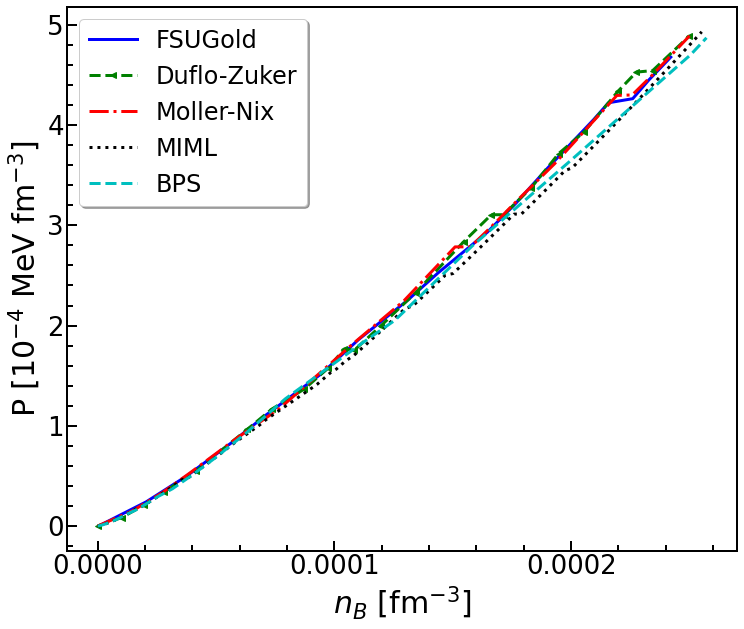}
\end{tabular}
\caption{The comparison of the pressure as a function of baryon density for the NS outer crust matter for different model.} 
\label{fig:press}
\end{figure}

{The results for the atomic composition of our NS outer crust EoS is in Figure \ref{fig:crustcompfig}. 
The upper panel shows the neutron number and lower panel shows the proton number of atomic nuclei populates 
in NS outer crust as a function of baryon density. As already mentioned in the introduction that NS outer crust 
can be classified into three regions. At the top layer of the outer crust (the density is low), the Coulomb 
lattice is populated by stable  $\mathrm{Fe}$ or $\mathrm{Ni}$ nuclei. As density increases, the electrons get captured on the protons and the system jumps to the energetically favorable $N = 50$ region. Finally, bottom layers of the outer crust are in the $N = 82$ 
region. In this region the theoretical extrapolations are unavoidable as almost no experimental information. We compare our results with a few representatives of two
classes of nuclear physics models. One class is that of microscopic/macroscopic models that yields $\sigma_{\rm rms}$ of the order of KeV while another class of models is based on accurately calibrated microscopic
properties of nuclei. We have chosen the Duflo and Zuker \cite{Duflo1995} and the finite range droplet model 
of Möller, Nix, and collaborators \cite{Moller1995} from the microscopic models and FSUGold \cite{ToddRutel2005} 
from calibrated models and BPS model from Baym, Pethick, and Sutherland \cite{Baym1971} for comparison. 
It is to be noted that in BPS model nuclear masses for the outer crust were provided by an early semi-empirical 
mass table. As can be seen from figure, our results are in very good agreement with the predictions of the other  
models. The composition profile is made of a sequence of plateaus. At the top layer of the NS outer crust in MIML, the Fe 
nuclei populate  upto $n_b= 5 \times 10^{-9}$ fm$^{-3}$ followed by the Ni (Z = 28) up to $n_b= 9.8 \times 10^{-7}$ fm$^{-3}$, 
before reaching the nucleus Kr (Z = 36) with N = 50. From this point,  nuclei face electron capture that reduces the proton number, 
resulting in a staircase structure upto Ni (Z = 28) at $n_b = 9 \times 10^{-5}$ fm$^{-3}$. Finally, the bottom layer
starts from 122Nb (Z = 41) all the way down to 116Kr (Z = 36), i.e, N = 80. Most physics models predict N = 82 
for Kr (Z = 36) at such densities whereas our MIML model predicts N =80. This seems to indicate the extrapolation 
limitation of MIML model \footnote{It is to be noted, the N = 82 plateau for Kr is far by N = 17 from
last available data 65Kr in AME2016}.

In Fig. \ref{fig:press} we compare the NS outer crust pressure of  the MIML model as a function of baryon density, with the four different models considered above. 
As can be seen from the figure that EoS obtained by MIML model is comparable to others. The entire data set for the MIML is given in Table \ref{tab:data} and can also be constructed from our codes
which are made publicly available.} It can be seen from Table \ref{tab:data} that for the larger densities,
several nuclei appear with a odd number of protons or neutrons. This seems to indicate another weakness of the MIML model which is not
 able to capture the pairing effect fully. This is in line with the results given in Table \ref{tab_rms} where the
$\sigma_{\rm rms}$ for \textit{even-even} nuclei is higher than the others. 

{As a last test, we have applied the MIML model to predict the BE  for the entire recently published  AME2020 data \cite{Huang:2021nwk} and for the  122 nuclei newly added \footnote{There are total 3557 ground state masses in AME2020 including 911 nuclei estimated from TMS. In this latest AME the mass precision is improved for 427 nuclei and masses of 74 nuclei is added experimentally in comparison with AME2016}. The $\sigma_{\rm rms}$ obtained in MIML for the entire AME2020 data is 0.2224 MeV, and for the newly added 122 is 0.7772 MeV. It is to be noted that the $\sigma_{\rm rms}$ of MIML is 0.58 MeV, see Table \ref{tab_out} (on the randomized test set). The $\sigma_{\rm rms}$ for the newly added 122 in AME2020 is slightly higher because of the few nuclei below Z = 20.}

\section{Conclusions} \label{sec5}
In keeping with the importance of BE in nuclear physics, 3435 nuclei have been measured (or estimated) in the laboratories around the world in the  atomic mass evaluation AME2016 \cite{Wang2017}.  {This table was recently updated with the publication of the AME2020 table \cite{Huang:2021nwk}.}
However a lot of excited and asymmetric neutron rich nuclei required in  astrophysical contexts cannot be produced in the lab. 
Theory also cannot come to the aid of experiments in these cases. The values of BE obtained within the present theories differ
from the experimentally observed values by $\sigma_{\rm rms}\approx 3$ MeV for the BW model \cite{Audi2012,Wang2012} to 
$\sigma_{\rm rms}\approx 0.3$ MeV for the  WS model \cite{Wang2014}. 
The error in the prediction in these theories is actually much larger for lighter nuclei and heavy nuclei with large neutron 
fraction which is of relevance for astrophysical studies.

In recent years, to overcome these problems in both theory and experiment, several attempts have been made to use ML algorithms on top of physics based models to predict the BE \cite{Gernoth1993,Athanassopoulos2004,Zhang2017,Utama2016,Niu2018}.
In these approaches, some physics model is used to calculate
the BE. The difference between this prediction and the experimentally observed values of BE forms the data set which is
then trained using ML algorithms. However, apart from the complexity of calculating the BE using some physics model, 
it is expected that these approaches will inherit some of the difficulties of the model itself. In this paper, we take
the next logical step on this road by eliminating the base physics model and explore the limitations of using only ML algorithms.
To the best of our knowledge, this
has not been attempted in literature before. We take the AME2016 data
\cite{Wang2017} and allow the ML algorithms to predict the BE. This we do via a two step process: the BE is calculated using some
ML algorithm and the error of this algorithm is further trained using another ML algorithm. We have obtained the best results
by using SVM as the base ML algorithm followed by 10 error training Random Forest algorithms. This model we denote as MIML model. 
A step by step exploration of various ML algorithms along with the justification of arriving at the best one is given in
the paper to help starting researchers to stop taking ML algorithms as a black box.

One of the key results of our work is that even with no physics inputs from theoretical nuclear physics models,
the MIML model has a $\sigma_{\rm rms}\approx 0.58$ on a randomized test set
which compares favorably with all the theoretical and ML improved theoretical models studied so far.
\footnote{Some of the data in AME2016 is obtained by TMS extrapolation, but that can considered to be
more of an \textit{experimental} fact than a result of theoretical modeling}.
The $\sigma_{\rm rms}$ obtained in
MIML model for light nuclei below $Z<20$ is 505 KeV and for nuclei between  $20\leq Z \leq 92$ is 248 KeV  while for heavy nuclei $Z>92$ it is 127 KeV.

The theoretical models actually give much larger $\sigma_{\rm rms}$ for lighter nuclei whereas the MIML model predicts BE with less $\sigma_{\rm rms}$ even for
lighter nuclei. This model may be extremely useful in determining BE of short lived
nuclei and has applications in constructing crusts of neutron stars. Our code for obtaining BE values for any nuclei of
choice is made publicly available for use in research purposes along with step by step instructions given in Appendix
\ref{programoverview}. This work can be taken as \textit{proof of concept}. 
It is possible that using more advanced Bayesian Neural Networks or Convolutional Neural Networks will further reduce the 
$\sigma_{\rm rms}$  {and will capture in a more efficient way the pairing effect and the identification of the magic numbers}. This will be attempted in a future work. Our MIML model can be used to predict other nuclear
ground state properties and compare them  with the predictions of well understood nuclear physics models as well as 
with experimental values given in AME2016. In a future work we  will attempt  to train ML models on other nuclear properties such as nuclear charge radii and
$\beta-$ decay half-lives.

Next we have constructed the outer crust of neutron stars using the predictions of BE from our MIML model. We show that our results are comparable to other existing models of the crust. The data set as well as the code is made publicly available for
verification and for use of the community. It is to be noted that the MIML model was constructed for interpolation and \textit{near} extrapolation. 
For nuclei far away from the AME2016 data our model may not give good predictions of the BE. However, some extrapolation is needed to construct the neutron
crust and the Gibbs free energy minimization algorithm we use. It has been constructed keeping in mind we should not be considering nuclei far away from
the AME2016 data set. The fact that we are able to get results comparable to other crust models indicate that our model is indeed capable of near
extrapolation. {We have seen that the model partially fails in performing a far extrapolation, in particular, if a far extrapolation is needed, magic numbers or pairing effects that favor even-even nuclei were not reproduced.} The present study  serves as a check for how far we can push the validity of ML model away from the experimental data.

Although in this work we show that we get good results in predicting the nuclear BE without introducing a nuclear physics model, {we have also identified weaknesses}. The role of theoretical models go way beyond producing numbers. A theoretical
model also indicates the actual physical mechanisms behind the properties being predicted. Since each term in the model
is physically motivated, a theoretical model which comes close to experimental predictions also identifies what are the actual
physical processes which are important in that energy scale.To have a theoretical understanding of any system, a physics based model is
necessary. ML algorithms cannot replace physics modeling in that respect.

\section*{ACKNOWLEDGMENTS} 
The authors would like to thank Prof. Snehanshu Saha for useful discussion and Prof J.N. De and Prof. B.K. Agarwal 
for a careful reading of the manuscript and for important suggestions.

\appendix
\section{Algorithm} \label{algo}
In this appendix we give a more technical formulation of the algorithm used. We start by choosing a base model out of LR, DT, RF, SVM, PR, and we train the model on our
training data. We tune the model parameters to avoid over-fitting and under-fitting using a 
validation set. We use the trained model to make predictions on the training data itself
and we store those predictions in {\bf current\_estimates} array. We initialize a variable {\bf count} with the value 1
which will keep track of how many Random Forest models are needed before the RMSE converges.
We also initialize two arrays {\bf error\_estimates} and {\bf error\_models}. Then we run a loop until the 
RMSE on validation set converges, and inside the loop we compute the difference in the actual 
data used for training {\bf (y\_train)} and the {\bf current\_estimates} array, and then train a Random Forest 
model to predict these differences, and we update the {\bf current\_estimates} array by adding the difference
to the previously predicted values. We also store the Random Forest models in an array to 
reuse them for testing. 

For testing, we first use the base model to make predictions on the test data, and then we use 
the {\bf count} number of Random Forest models to make prediction about the difference in those values.
We the add those differences to the prediction to get an improved prediction.
A schematic description of the algorithm is given in Algorithm \ref{algo1}.

\begin{figure*}[htp]
\begin{minipage}{\linewidth}
\begin{algorithm}[H]
\caption{\label{algo1} Error Training and Testing} 
\begin{algorithmic}[1]
	\State //Training base model
	\State base\_model.fit(X\_train, y\_train)
	\State current\_estimates $\leftarrow$ base\_model.predict(X\_train)
	\State count $\leftarrow 1$
	\State error\_estimates $\leftarrow []$
	\State error\_models $\leftarrow []$
	\State //Repeated Error training and saving models in error\_models array
		\While { RMSE on validation set not converged}
		    \State error $\leftarrow$ y\_train - current\_estimates
		    \State error\_models[count] $\leftarrow$ RandomForest.fit(X\_train, error)
		    \State error\_estimates $\leftarrow$ error\_models[count].predict(X\_train)
		    \State current\_estimates $\leftarrow$ current\_estimates + error\_estimates
		    \State count $\leftarrow$ count+1
		    \State Compute RMSE on validation set
		\EndWhile
		
    \State //Testing on Test data
    \State test\_estimates $\leftarrow$ base\_model.predict(X\_train)
    \For {$i = 1, 2...count$}
        \State error\_estimates\_test $\leftarrow$ error\_models[i].predict(X\_test)
        \State test\_estimates $\leftarrow$ test\_estimates + error\_estimates\_test
    \EndFor
    \State Final Prediction on Test Set is test\_estimates
	\end{algorithmic} 
\end{algorithm}
\end{minipage}
\end{figure*}

\section{Program overview} \label{programoverview}
\subsection{Program to run MIML Model}
The best algorithm of this work (MIML) is made publicly available via GitHub so that the  physics community is able to use to 
different applications. The summarized algorithm is given in Appendix A. The details of the program is as follows.

{\bf Installation--}
The repository containing the trained models is uploaded on GitHub link : \href{https://github.com/be-prediction-bitsgoa/nuclear-mass-prediction}{\bf (https://github.com/be-prediction-bitsgoa/nuclear-mass-prediction)}. To get the model running, click on the link provided and clone the repository. {Note that since the files are large, Git LFS must be initialize before cloning}
\\
Alternatively, the files can also be downloaded from Google Drive link: 
\href{https://drive.google.com/drive/folders/1EWUot993Ci-BxP99V8DFrNbzKTuLLLaf}{\bf (https://drive.google.com/drive/folders/1EWUot993Ci-BxP99V8DFrNbzKTuLLLaf)}. Move into the folder containing the following files:

\begin{enumerate}
    \item base\_model.sav
    \item error\_model\_1.sav
    \item error\_model\_2.sav
    \item error\_model\_3.sav
    \item error\_model\_4.sav
    \item error\_model\_5.sav
    \item error\_model\_6.sav
    \item error\_model\_7.sav
    \item error\_model\_1.sav
    \item error\_model\_8.sav
    \item error\_model\_9.sav
    \item error\_model\_10.sav
    \item driver.py
    \item requirements.txt
\end{enumerate}
The .sav files are the trained models in pickled form, and driver.py is the python script which will run these models to make prediction. 
The program has a few dependencies that need to be installed before running the script.
\begin{enumerate}
\item System must have python3 (64-bit) to run driver.py script.
\item System must have pip downloaded. Pip is a package manager for python and it is required to facilitate installation of other python libraries like NumPy, Pandas, Scikit-learn etc. 
\item Open terminal (for ubuntu) or command prompt (for windows) in the directory containing the above mentioned files.
\item Install all the required libraries by invoking the following command in terminal/command prompt: \\
pip install -r requirements.txt
\end{enumerate}

{\bf Prediction--}
To make prediction using the models, a csv file needs to be created in the same folder as the driver.py file. This csv file will contain the Z and N values 
of the nuclei for which predictions are to be made. The csv file must contain 2 columns, first one for Z and second one for N. The columns should not have any headers. 
Save and close this csv file before running driver.py.

Now, run driver.py through command line by typing "py driver.py" for windows or "python driver.py" for ubuntu systems. Upon running, there will be a prompt 
asking for the name of the csv file where the test data is kept. Enter the name of the file along with the .csv extension and press Enter. The binding energy predicted by the models for the
 nuclei specified in the csv file will be displayed on the terminal/command prompt in tabular form. 
 
\subsection{Neutron Start outer crust EoS program}
The program for determining the composition of the outer crust of neutron star using the MIML model can also be found in the same drive link mentioned in part 1, in the folder named "Crust". The folder "Crust" consists of 3 files : 
\begin{enumerate}
    \item crust.py
    \item data\_precalculated.csv
    \item nz\_range.csv
\end{enumerate}

crust.py consists of the main python code to estimate the composition of outer crust of neutron stars over a range of
densities using the predictions of the MIML model. The other 2 files are auxiliary files which are used by crust.py
while running, so they must be closed before running crust.py. data\_precalculated.csv consists of MIML model
predictions for a large number of Z and N combinations, which are precalculated and stored to make the code faster.
nz\_range.csv consists of Z and N numbers available in the AME2016 data set, and it is used to compute the permissible range of Z,N combinations such that we are not extrapolating very far from the AME2016 data set region.

{\bf Running the code--}
To run the code:
\begin{enumerate}
    \item first make sure that all the 3 mentioned files are in the same directory
    \item Open terminal (for Ubuntu) or command prompt (for Windows) in the directory containing the above mentioned files.
    \item Run crust.py by typing "py crust.py" for windows or "python crust.py" for ubuntu
    \item After completion, the output will be saved in a file called output\_final.csv.
\end{enumerate}
\pagebreak

\section{Table for Composition and Equation of State for Neutron Star Crust} \label{datatable}
{\setlength{\tabcolsep}{5pt}
\setlength{\tabulinesep}{1.4pt}
\begin{longtable}{cccccc}
\caption{Composition and equation of state of the outer crust} \label{tab:data}\\
\hline 
$n_{\rm B}$     & Z  & A  & BE  & $\varepsilon$ &  $P$      \\ 
fm$^{-3}$  & & & MeV & MeV fm$^{-3}$ & MeV fm$^{-3}$ \\ \hline \hline 
\endhead
\hline
\endfoot
\endlastfoot
8.4021E-12  & 26 & 56  & 492.257 & 7.8155E-09      & 1.0190E-14      \\
7.7376E-11  & 26 & 56  & 492.257 & 7.1974E-08      & 4.6931E-13      \\
7.4879E-10  & 26 & 56  & 492.257 & 6.9653E-07      & 1.9039E-11      \\
5.8253E-09  & 28 & 62  & 545.258 & 5.4192E-06      & 4.1077E-10      \\
7.2002E-09  & 28 & 62  & 545.258 & 6.6982E-06      & 5.5941E-10      \\
2.8613E-08  & 28 & 62  & 545.258 & 2.6620E-05      & 3.9790E-09      \\
9.0416E-08  & 28 & 62  & 545.258 & 8.4127E-05      & 1.9466E-08      \\
1.8019E-07  & 28 & 62  & 545.258 & 1.6767E-04      & 4.9756E-08      \\
1.8562E-07  & 26 & 58  & 509.948 & 1.7273E-04      & 5.1393E-08      \\
1.9606E-07  & 28 & 64  & 561.752 & 1.8245E-04      & 5.3437E-08      \\
3.4870E-07  & 28 & 64  & 561.752 & 3.2451E-04      & 1.1636E-07      \\
5.5260E-07  & 28 & 64  & 561.752 & 5.1429E-04      & 2.1631E-07      \\
8.7579E-07  & 28 & 64  & 561.752 & 8.1515E-04      & 4.0155E-07      \\
9.0979E-07  & 28 & 66  & 576.800 & 8.4683E-04      & 4.0551E-07      \\
9.8618E-07  & 36 & 86  & 749.228 & 9.1799E-04      & 4.4022E-07      \\
1.0486E-06  & 36 & 86  & 749.228 & 9.7608E-04      & 4.7797E-07      \\
1.3198E-06  & 36 & 86  & 749.228 & 1.2287E-03      & 6.5069E-07      \\
1.6615E-06  & 36 & 86  & 749.228 & 1.5467E-03      & 8.8573E-07      \\
1.9490E-06  & 36 & 86  & 749.228 & 1.8145E-03      & 1.0968E-06      \\
2.0259E-06  & 34 & 84  & 727.332 & 1.8862E-03      & 1.1064E-06      \\
3.2106E-06  & 34 & 84  & 727.332 & 2.9895E-03      & 2.0486E-06      \\
5.0885E-06  & 34 & 84  & 727.332 & 4.7388E-03      & 3.7917E-06      \\
6.8899E-06  & 34 & 84  & 727.332 & 6.4170E-03      & 5.6843E-06      \\
7.4360E-06  & 32 & 82  & 702.223 & 6.9264E-03      & 6.0063E-06      \\
9.8033E-06  & 32 & 82  & 702.223 & 9.1323E-03      & 8.6877E-06      \\
1.2342E-05  & 32 & 82  & 702.223 & 1.1498E-02      & 1.1815E-05      \\
1.6269E-05  & 32 & 82  & 702.223 & 1.5159E-02      & 1.7084E-05      \\
1.7976E-05  & 30 & 80  & 673.877 & 1.6752E-02      & 1.8545E-05      \\
2.3699E-05  & 30 & 80  & 673.877 & 2.2087E-02      & 2.6816E-05      \\
2.9834E-05  & 30 & 80  & 673.877 & 2.7809E-02      & 3.6460E-05      \\
3.5827E-05  & 30 & 80  & 673.877 & 3.3398E-02      & 4.6547E-05      \\
3.7707E-05  & 28 & 78  & 641.541 & 3.5156E-02      & 4.7109E-05      \\
5.3147E-05  & 28 & 78  & 641.541 & 4.9561E-02      & 7.4467E-05      \\
5.6693E-05  & 28 & 78  & 641.541 & 5.2869E-02      & 8.1168E-05      \\
6.8476E-05  & 28 & 80  & 646.396 & 6.3872E-02      & 1.0095E-04      \\
7.1079E-05  & 28 & 80  & 646.396 & 6.6301E-02      & 1.0610E-04      \\
9.2623E-05  & 41 & 122 & 948.654 & 8.6439E-02      & 1.4126E-04      \\
9.8448E-05  & 41 & 122 & 948.654 & 9.1878E-02      & 1.5323E-04      \\
1.0083E-04  & 40 & 120 & 927.590 & 9.4105E-02      & 1.5663E-04      \\
1.0459E-04  & 40 & 120 & 927.590 & 9.7613E-02      & 1.6446E-04      \\
1.0830E-04  & 40 & 121 & 929.054 & 1.0108E-01      & 1.7040E-04      \\
1.1197E-04  & 40 & 121 & 929.054 & 1.0451E-01      & 1.7814E-04      \\
1.1560E-04  & 40 & 121 & 929.054 & 1.0790E-01      & 1.8589E-04      \\
1.1919E-04  & 40 & 121 & 929.054 & 1.1126E-01      & 1.9363E-04      \\
1.2275E-04  & 40 & 121 & 929.054 & 1.1458E-01      & 2.0138E-04      \\
1.2627E-04  & 40 & 121 & 929.054 & 1.1787E-01      & 2.0912E-04      \\
1.2976E-04  & 40 & 121 & 929.054 & 1.2113E-01      & 2.1687E-04      \\
1.3322E-04  & 39 & 120 & 908.364 & 1.2438E-01      & 2.1978E-04      \\
1.3665E-04  & 39 & 120 & 908.364 & 1.2758E-01      & 2.2736E-04      \\
1.4005E-04  & 39 & 120 & 908.364 & 1.3076E-01      & 2.3494E-04      \\
1.4343E-04  & 39 & 120 & 908.364 & 1.3391E-01      & 2.4252E-04      \\
1.4678E-04  & 39 & 120 & 908.364 & 1.3704E-01      & 2.5010E-04      \\
1.5010E-04  & 38 & 119 & 887.074 & 1.4015E-01      & 2.5193E-04      \\
1.5340E-04  & 38 & 119 & 887.074 & 1.4323E-01      & 2.5934E-04      \\
1.5667E-04  & 38 & 119 & 887.074 & 1.4629E-01      & 2.6675E-04      \\
1.5992E-04  & 38 & 119 & 887.074 & 1.4933E-01      & 2.7416E-04      \\
1.6315E-04  & 38 & 119 & 887.074 & 1.5235E-01      & 2.8157E-04      \\
1.6636E-04  & 38 & 119 & 887.074 & 1.5535E-01      & 2.8898E-04      \\
1.6955E-04  & 38 & 119 & 887.074 & 1.5833E-01      & 2.9639E-04      \\
1.7272E-04  & 38 & 119 & 887.074 & 1.6129E-01      & 3.0380E-04      \\
1.7587E-04  & 38 & 119 & 887.074 & 1.6423E-01      & 3.1121E-04      \\
1.7900E-04  & 37 & 118 & 864.810 & 1.6717E-01      & 3.1127E-04      \\
1.8211E-04  & 37 & 118 & 864.810 & 1.7008E-01      & 3.1851E-04      \\
1.8521E-04  & 37 & 118 & 864.810 & 1.7298E-01      & 3.2574E-04      \\
1.8828E-04  & 37 & 118 & 864.810 & 1.7585E-01      & 3.3298E-04      \\
1.9134E-04  & 37 & 118 & 864.810 & 1.7871E-01      & 3.4022E-04      \\
1.9439E-04  & 37 & 118 & 864.810 & 1.8156E-01      & 3.4746E-04      \\
1.9742E-04  & 37 & 118 & 864.810 & 1.8439E-01      & 3.5470E-04      \\
2.0043E-04  & 36 & 116 & 841.992 & 1.8722E-01      & 3.5734E-04      \\
2.0343E-04  & 36 & 116 & 841.992 & 1.9002E-01      & 3.6449E-04      \\
2.0641E-04  & 36 & 116 & 841.992 & 1.9281E-01      & 3.7164E-04      \\
2.0938E-04  & 36 & 116 & 841.992 & 1.9558E-01      & 3.7878E-04      \\
2.1234E-04  & 36 & 116 & 841.992 & 1.9835E-01      & 3.8593E-04      \\
2.1528E-04  & 36 & 116 & 841.992 & 2.0110E-01      & 3.9308E-04      \\
2.1821E-04  & 36 & 116 & 841.992 & 2.0383E-01      & 4.0023E-04      \\
2.2112E-04  & 36 & 116 & 841.992 & 2.0656E-01      & 4.0737E-04      \\
2.2403E-04  & 36 & 116 & 841.992 & 2.0927E-01      & 4.1452E-04      \\
2.2692E-04  & 36 & 116 & 841.992 & 2.1197E-01      & 4.2167E-04      \\
2.2980E-04  & 36 & 116 & 841.992 & 2.1467E-01      & 4.2881E-04      \\
2.3266E-04  & 36 & 116 & 841.992 & 2.1734E-01      & 4.3596E-04      \\
2.3552E-04  & 36 & 116 & 841.992 & 2.2001E-01      & 4.4311E-04      \\
2.3836E-04  & 36 & 116 & 841.992 & 2.2267E-01      & 4.5026E-04      \\
2.4119E-04  & 36 & 116 & 841.992 & 2.2532E-01      & 4.5740E-04      \\
2.4401E-04  & 36 & 116 & 841.992 & 2.2796E-01      & 4.6455E-04      \\
2.4682E-04  & 36 & 116 & 841.992 & 2.3058E-01      & 4.7170E-04      \\
2.4962E-04  & 36 & 116 & 841.992 & 2.3320E-01      & 4.7884E-04      \\
2.5241E-04  & 36 & 116 & 841.992 & 2.3581E-01      & 4.8599E-04      \\
2.5519E-04  & 36 & 116 & 841.992 & 2.3840E-01      & 4.9314E-04      \\ \hline
\end{longtable}}

\bibliography{biblio}

\end{document}